\documentclass[prb,superscriptaddress,nobalancelastpage,twocolumn, showpacs, notitlepage]{revtex4-1}
\usepackage[normalem]{ulem}
\usepackage{cancel}
\usepackage{graphicx}
\usepackage{amssymb}
\usepackage{natbib}
\usepackage{dcolumn}
\usepackage{bm}
\usepackage[caption=false]{subfig}
\usepackage{graphicx,psfrag,xspace}
\usepackage{xcolor}
\usepackage{amssymb,amsfonts,amsmath}
\usepackage{setspace}
\usepackage{mathrsfs}
\DeclareMathAlphabet{\mathpzc}{OT1}{pzc}{m}{it}
\def\bra#1{\mathinner{\langle{#1}|}} 
\def\ket#1{\mathinner{|{#1}\rangle}} 
\def\braket#1{\mathinner{\langle |{#1}|\rangle}}

\def\omw{\omega_{\text{\tiny MW}}}
\usepackage{comment}

\newcommand{\tinytext}[1]{\mbox{\tiny #1}}

\newcommand{\tmax}{t_{\mbox{\tiny max}}}

\newcommand{\HH}[1]{\hat{H}_{\text{\tiny #1}}}

\def\Hmw{\HH{MW}}

\usepackage{empheq}

\newcommand{\citeo}[1]{[\onlinecite{#1}]}

\newcommand{\titleinfo}{An exactly solvable model for dynamic nuclear polarization}

\newcommand{\Tr}{\operatorname{Tr}}

\def\pnstat{p_n^{\mbox{\tiny stat}}}

\newcommand{\citei}[1]{[\onlinecite{#1}]}

\graphicspath{{figures/},{figures/final/}}

\begin{document}

\author{In\'es Rodr\'iguez-Arias}
\affiliation{LPTMS, CNRS, Univ. Paris-Sud, Universit\'e Paris-Saclay, 91405 Orsay, France}
\author{Markus M\"uller}
\affiliation{Condensed matter theory, Paul Scherrer Institute, CH-5232 Villigen PSI, Switzerland}
\affiliation{The Abdus Salam International Centre for Theoretical Physics, 34151, Trieste, Italy} 
\author{Alberto Rosso}
\affiliation{LPTMS, CNRS, Univ. Paris-Sud, Universit\'e Paris-Saclay, 91405 Orsay, France}
\author{Andrea De Luca}
\affiliation{The Rudolf Peierls Centre for Theoretical Physics, Oxford University, Oxford, OX1 3NP, United Kingdom}

\title{\titleinfo}
\begin{abstract}

We introduce a solvable model of driven fermions that elucidates the role of the localization transition in driven disordered magnets, as used in the context of dynamic nuclear polarization. 
Instead of spins, we study a set of non-interacting fermions that are coupled locally to nuclear spins and tend to hyperpolarize them. The induced hyperpolarization is a fingerprint of the driven steady state of the fermions, which undergo an Anderson Localization (AL) transition upon increasing the disorder. Our central result is that the maximal hyperpolarization level is always found close to the localization transition. In the limit of small nuclear moments the maximum is pinned to the transition, and the hyperpolarization is strongly enhanced by multi-fractal correlations in the critical state of the nearly localized driven system, its magnitude reflecting multi-fractal scaling.
\end{abstract}

\maketitle 
\section{Introduction}
Statistical mechanics is grounded on the assumption that simple macroscopic laws emerge whenever time evolution is so complex
that specific details become irrelevant. Although this picture is generically confirmed in the evolution of closed quantum systems,
recent work\cite{basko2006metal} has predicted that strong disorder can prevent 
thermal equilibrium when a quantum system evolves under its own dynamics. 
Under certain circumstances (low dimension lattice models, short-range interactions), this phenomenon occurs as a sharp transition which goes under the name of
many-body localization (MBL). In higher dimensions or in systems with power-law interactions, the MBL transition is believed to become a sharp crossover. 
Experimental measurements of MBL have been reported 
in highly controlled experimental settings, such as ultra-cold  atomic
fermions,\cite{schreiber2015observation, bordia2016coupling} and chains of trapped ions.\cite{smith2016many}
Most theoretical studies have so far focused on isolated one-dimensional lattice systems which display the main features of the MBL
phase.\cite{pal2010many, Vosk2013a, serbyn2013local, de2013ergodicity, luitz2015many, ros2015integrals, pino2016nonergodic, imbrie2016many}

However, a small coupling to the environment is present in any physical system, even in cold atoms and trapped ions, while standard solid state systems always feel the presence of the bath of phonons on sufficiently long time scales.
For this reason,
recent studies have started to address the fate of the MBL  transition in the presence of 
weak dissipation.\cite{Nandkishore2014, fischer2016dynamics, levi2016robustness}
On sufficiently long time-scales, the presence of a non-localized environment always restores ergodicity in an otherwise localized phase. 
Nevertheless, when a weakly dissipative system is driven out-of-equilibrium, it can reach a stationary state
whose characteristics will strongly depend on whether its intrinsic dynamics is ergodic or non-ergodic 
-- even though, no genuine transition survives in the space of steady states, but only a strong crossover~\cite{deluca2016,rosch2018}. 
Up to date, mainly two experimental settings 
of driven dissipative systems have been studied in this context: interacting gases of Rydberg atoms~\cite{marcuzzi2016localization} 
and dynamic nuclear polarization (DNP) protocols.\cite{DeLuca2015, deluca2016, caracciolo2016evidence} Here we focus on the latter, 
for which it has been shown that the MBL crossover of the closed system manifests itself in key features of the driven-dissipative
steady state.

DNP is a very promising technique to cool nuclear spins and thereby improve the signal-to-noise ratio in nuclear magnetic 
resonance (NMR) measurements. The procedure is as follows: a glassy (amorphous) sample of the nuclear compound is doped with radicals, i.e.,
molecules with unpaired electron spins. The sample is then cooled down to temperatures of the order of $1$~K and 
subjected to a strong magnetic field $B$, which in many standard experiments is fixed to $B=3.35$~T.
Under these conditions, electron spins have a high polarization level, whereas nuclear spins remain almost 
unpolarized and thus yield a very small NMR signal. 
The sample is then driven out of equilibrium by a microwave field. 
By tuning the frequency of the microwaves 
close to the Zeeman gap of the electron spins, one observes a huge transfer of polarization
from the electron spin system to the nuclear one, which results in a strongly enhanced NMR contrast.  In order to optimize the resulting cooling, it is thus paramount to understand the underlying polarization transfer mechanism.

In a recent work,\cite{deluca2016} we showed that the efficiency of the procedure (i.e., the final nuclear polarization)
depends on the interplay between the disorder arising from the $g$-factor anisotropy of the electron spins,
and their dipolar interactions, whose magnitude is tuned in practice with the radical concentration. For strong interactions, the electron spins are ergodic, and their  stationary state can be described by an effective spin temperature.\cite{Borghini1968, Abragam1982a}
In contrast, when disorder is dominant and the closed system is localized, the stationary nuclear polarization was found to rapidly drop.
As a consequence, the optimal value is attained close to the MBL crossover of the isolated system.

These results were obtained by exactly diagonalizing an interacting electron system,
which, however, was limited to small system sizes with at most $13$ spins. 
Moreover, that study could not furnish a thorough understanding of the crossover from the ergodic into the localized regime, and the associated maximum in the cooling effect. 
In order to go beyond these limitations, here we present the study of a free-fermion analogue of the electron spin problem,
which allows us to much larger sizes ($\simeq 10^4$). Because of integrability, the
free fermions do not have a genuine ergodic phase. Nevertheless,
disorder tunes a single-particle Anderson localization (AL) transition, which we find to result in the same key parameters of the driven steady state as in the many-body spin system. In particular,  the maximal nuclear polarization is reached
when the electrons are close to the localization transition. Our toy model has the crucial advantage that the properties of this crossover are much better accessible to analytical insight.
Indeed, the stationary nuclear polarization can be put in direct relation to 
basic quantities which describe the critical behavior of the Anderson transition, 
such as the inverse participation ratio and the (multi-fractal) eigenfunction correlations.

In Sec.~\ref{DNPProtocol} we present the DNP protocol and some review of the theoretical treatments studied so far. In Sec.~\ref{Sec:WeakInt}, we present an approximation reasonable for weak interactions that allows exact computations of the nuclear polarization. In Sec.~\ref{Sec:Results} we compute the nuclear polarization and present it in function of the strength of the electron dipolar interactions. Finally, in Sec.~\ref{Sec:Discussion} we discuss its behavior and link it to the localization/ergodicity properties of the electron spin system. 

\section{Microscopic description of DNP\label{DNPProtocol}}

The DNP protocol operates with two spin species: nuclear and electronic. 
Intra-species interactions are given by dipolar couplings, while nuclear and electron spins
are coupled with each other through hyperfine interactions. The Hamiltonian of the isolated spin system can thus 
be written as the sum of three terms:
\begin{equation}
\label{HamSpin}
\HH{S}=\HH{e}+\HH{n}+\HH{e--n}\, ,
\end{equation}
where the first and second terms stand, respectively, for the electronic and nuclear subsystems and the last term describes the hyperfine interactions. 
Let us describe how we treat each term in more detail:
\begin{subequations}
\begin{align}
\label{dipham}
\HH{e} &= \sum_{j = 1}^{N_e} \omega_j \hat{S}_j^z +
\HH{e}^{(dip)}
\\
 \HH{n} &= -\omega_n \sum_{p= 1}^{N_n} \hat{I}_p^z + 
\HH{n}^{(dip)}
 \;,
\\
 \HH{e--n} &= \sum_{j=1}^{N_e} A_{j} \hat{S}_j^z \hat{I}_j^x \;,
\label{hHF}
\end{align}
\end{subequations}
where $\hat{S}_j^\ell$ and $\hat{I}_p^\ell$ are respectively the $\ell$-component corresponding to the $N_e$ electronic and $N_n$ nuclear spins (with $N_n\gg N_e$). The Larmor frequencies $\omega_j$ and $\omega_n$ are proportional to the external magnetic field. 
Electron spins display inhomogeneities in the Larmor frequencies, i.e. $\omega_j = \omega_e + \delta\omega_j$, due to 
the local $g$-factor anisotropy, while such anisotropies are negligible for the nuclear spins. 
The coefficients 
$A_{j}$
indicate the strength of 
hyperfine interactions: here, we assume that, among the $N_n$ nuclear spins, only $N_e\ll N_n$ are ``{\it core}'' nuclear spins, 
i.e. they have a significant interaction with the electron spins. For the sake of simplicity, 
we assume that each core nucleus is attached to one and only one electron spin. We also assume that there are no electron spins without a ``{\it core}'' nucleus. 
We use $\hat I_j^\ell$, $j=1,\dots, N_e$, to label those spins and  $\hat I_p^\ell$, $p=1,\dots,N_n$, to label the full set of spins. 
The precise form of the dipolar Hamiltonians $\HH{e/n}^{(dip)}$ is given in Appendix \ref{DiagHam}.

The spin system is weakly coupled to a (phonon) reservoir at a temperature $\beta^{-1}$. It is irradiated by a microwave field 
of frequency $\omw$ that couples with an amplitude $\omega_1$ to the electron spins, via a Hamiltonian  $\Hmw = \omega_1 \sum_j \hat{S}^x_j \cos(\omw t)$. 

The full Hamiltonian of the system reads :
\begin{align}
\label{hTotal}
 \HH{total} =   \HH{e} + \HH{n} + \HH{e--n}  + \HH{R} + \HH{S--R} + \HH{MW} \;, 
\end{align}
The term  $\HH{R}$ governs the dynamics of the reservoir and $\HH{S--R}$ contains the coupling of the spin system with the reservoir. 
It would  be a formidable task to describe this system exactly,
and one is thus forced to resort to certain approximations. 

\subsection{Review of Lindblad approximation schemes}
A natural approach is to treat separately fast and slow 
time-scales in the full system: $\HH{total} = \HH{fast} + \HH{slow}$, where $\HH{fast}$ and $\HH{slow}$ will be specified later on. 
This allows us to perform the weak-coupling approximation, leading to an equation for the density
matrix of the spin system in a Lindblad form:\cite{petruccione2002theory}
the fast unitary dynamics due to $\HH{fast}$ is treated exactly, while
the contribution of $\HH{slow}$ is taken into account perturbatively.
Within this approximation, one can 
obtain a further simplification by projecting the Lindblad equation 
onto the eigenstates of $\HH{fast}$. In this basis, the density matrix of the spin system is always diagonal and the dynamics reduces to a classical master equation describing the evolution of
the probabilities of occupation of the eigenstates $\ket{n}$ of $\HH{fast}$: $\HH{slow}$ 
only determines the transition rates between pairs of eigenstates $\ket{n}, \ket{m}$ 
through matrix elements of all the local operators $\hat O$ involved in $\HH{slow}$, i.e. $|\bra{n} \hat{O} \ket{m}|^2$. 
Within this approach, to take into account the time-dependent microwaves, 
one resorts to the rotating-wave approximation,~\cite{Abragam1982a}
which amounts to neglecting rapidly oscillating terms and thereby obtaining a static Hamiltonian in the rotating frame.

Different schemes are obtained according to how $\HH{total}$ is split into $\HH{fast}$ and $\HH{slow}$.
Given that the coupling with the reservoir and microwaves is always weak,
a first possibility, discussed in [\onlinecite{deluca2016},\onlinecite{Hovav2012},\onlinecite{Hovav2013}],
is to include in $\HH{slow}$ the last three terms of \eqref{hTotal}.
Although this already represents a significant simplification of the original problem, one still faces the non-trivial task 
of diagonalizing the interacting Hamiltonian in \eqref{HamSpin} 
and computing the appropriate matrix elements between its eigenstates. 
In the absence of further simplifications, one is limited to the use of exact diagonalization
and small system sizes (maximum of $20$ electron spins). This  route was followed in [\onlinecite{deluca2016}]. 

A second approach \cite{karabanov2012quantum,InesSolidEffect,karabanov2015dynamic}
consists in including into $\HH{slow}$ not only the coupling with the reservoir and the microwaves, but also 
all the interaction terms inside $\HH{S}$. In this way, 
the eigenstates $\ket{n}$ become simply product states of spins polarized along the direction of the magnetic field and, 
all the transition rates can be computed analytically. The solution of the master equation can be obtained 
with Monte Carlo simulations for much larger system sizes. However, in order for this approach to be efficient, 
all the rates have to be of the same order of magnitude: this was successfully used in 
[\onlinecite{karabanov2012quantum}, \onlinecite{karabanov2015dynamic}] 
in the presence of a single electron spin to model 
the process of nuclear spin diffusion. Instead, when many electron spins are considered,
the presence of inhomogeneities and their fast dynamics 
render this method inefficient. 
Moreover, in the purpose of the present paper, we cannot entirely follow this approach: indeed the product states have no mixing and thus they do not capture the ergodicity properties of the DNP system. Nonetheless, inspired by this approach, here we include the hyperfine interactions into $\HH{slow}$,\cite{InesSolidEffect} which is already an important simplification and accounts for the ergodicity properties of the electron spin system, found to be the most relevant to study the MBL crossover.~\cite{deluca2016}

\section{DNP in the limit of weak dipolar interactions \label{Sec:WeakInt}}
Here, we propose a different approach that allows us to deal with a large number of degrees of freedom ($N_e$ up to $10^4$).
We treat perturbatively the hyperfine interaction $\HH{e--n}$, so that the eigenstates of the whole spin system
factorize as
\begin{equation}
\label{factorization}
 \ket{n} = \ket{A} \otimes \ket{\mu}
\end{equation}
with $\ket{A}$ an eigenstate of $\HH{n}$ and $\ket{\mu}$ one of $\HH{e}$. For the nuclear system,
given the negligible inhomogeneity of the Larmor frequencies,
it is treated as a perfectly ergodic many-body system, which satisfies the eigenstate thermalization hypothesis (ETH).

For the electron spins, in order to treat the competition between dipolar interactions and inhomogeneous Larmor frequencies,
we resort to a strong simplification.

\subsection{Simplified model for the electron spins: the free fermion approximation}
Let us replace the bosonic spin degrees of freedom with spin-less fermions
\begin{align}
\label{replSferm}
\hat{S}_j^x \to \frac{\hat{c}_j^\dag + \hat{c}_j}{2}\;, \,
\hat{S}_j^y \to -\imath\frac{\hat{c}_j^\dag - \hat{c}_j}{2}\;,\,
\hat{S}_j^z \to \frac{1}{2} - \hat{c}_j^\dag \hat{c}_j\;. 
\end{align}
so that a value of $S_j^z=\pm 1/2$ corresponds to an empty or occupied site $j$, respectively.
The fermionic operators satisfy the anti-commutation relation $\{ \hat{c}_i^\dag, \hat{c}_j\} = \delta_{ij}$. 
For a single-site $j$, this is an exact mapping, as it correctly reproduces all the spin commutation relations. 
However, for different sites, spins commute while fermions anti-commute. 
This problem can be solved by a non-local Jordan-Wigner tail to the fermions (upon defining an order of the sites). In higher dimensions, while the introduction of meandering Jordan-Wigner tails is in principle possible,
such an exact mapping becomes impractical and one is forced to make some approximations.
Here, we simply drop the Jordan-Wigner tail in order to obtain a tractable model and make the substitutions \eqref{replSferm} in \eqref{dipham}.
Since we are interested in  qualitative aspects of the model, we simplify even further by dropping interaction terms, and restricting fermion hoppings to nearest neighbors to obtain the following Hamiltonian
(see Appendix \ref{DiagHam} for a detailed discussion):
\begin{equation}
\HH{A} =\sum_i \Delta_i c_i^\dagger c_i - t \sum_{\langle i,j \rangle}  \left( 
c_i^\dagger c_j+ c_j^\dagger c_i \right) + \omega_e \sum_i c_i^\dagger c_i,
\label{hAnd}
\end{equation}
where the $\Delta_i$ are uniformly distributed in the interval $[- w /2, w/2]$. 
The hopping parameter $t$ between nearest neighbors
parameterizes the strength of dipolar interaction,
as for instance, after our replacement \eqref{replSferm} $\hat S_j^+ \hat S_k^- \to c_j^\dagger c_k$. 
Under standard DNP conditions for trityl radicals, the external magnetic field of $B=3.35$~T is responsible for the ``\emph{chemical potential}'' term with $\omega_e = 93.9\times2\pi$~GHz and, 
via the g-factor anisotropy, for the strength of the disorder characterized by the width of its distribution, $w \sim 108 \times2\pi$~MHz. 
The Hamiltonian~\eqref{hAnd} is simply the $3D$-Anderson model. Its 
main advantage is that it combines
solvability with the presence of a localization transition 
at a critical value of the ratio between disorder and hopping, which for a cubic array of sites and box-distributed disorder takes the value $w/t=(16.536\pm 0.007)$ \citeo{slevin2014}.
This allows us to study qualitative features of the localization transition and their effect on the driven steady state. Note that the algebraic tail of the dipolar interactions would translate into algebraic hoppings, which are known to lead to delocalization even in strong disorder.\cite{Anderson1958} However, the associated dynamic time scale just  grows exponentially with disorder strength instead of truly diverging at a critical disorder. Once this dynamic time scale exceeds the long time scales associated with the driving and heat bath coupling, it becomes irrelevant whether the internal dynamics  is genuinely localized or simply very slow. For that reason the algebraic tails in the hopping do not seem crucial to understand the properties of the steady state at the disorder strength where the two time scales become comparable. We therefore restrict ourselves to a short range model with a sharp localization transition. This allows us to take the limit of infinitely weak coupling to the drive and the bath, and thereby obtain useful theoretical insight into the effects of incipient localization  on the driven steady state.

Before we proceed to analyze the driven dynamics of  
this free-fermion model, let us comment about the status of \eqref{hAnd} and its validity in describing the original spin system. Let us point out that we are not interested in a quantitatively accurate description of the behavior of the driven spin system. Our aim is rather to understand how qualitative features of the localization transition enter into properties of the steady state. 
In the extremely localized limit ($t\to 0$), i.e. when the disorder dominates over the dipolar interactions, the fermions correctly describe decoupled electron spins. For weak dipolar interactions, a small hopping term can capture qualitatively the perturbative regime. However, increasing the dipolar interactions, the spins become more and more interacting and undergo increasingly strong quantum fluctuations. The corresponding increase of the hopping term in \eqref{hAnd} also leads to a larger localization length, however still within a system of non-interacting fermions. This mere delocalization effect is of course not equivalent to the more complex interaction and fluctuation effects in the spin model, as \eqref{hAnd} can never describe an ergodic thermal reservoir. Nevertheless, we take the fermionic model as a useful, simplified but tractable toy model in which to investigate the role played by the Anderson localization transition and evaluating analogs of the correlation functions that are relevant in the driven spin system. The results from the fermionic study provide qualitative insight on the crossover from MBL to ergodicity occurring in the original spin model of Eq.~\eqref{dipham}.

\subsection{Diagonalization of the free-fermion Hamiltonian}
To construct the eigenstates of \eqref{hAnd}, one introduces the fermionic operators $a_{\alpha}=\sum_i \phi^*_{\alpha i}  c_i$, 
or equivalently $c_i=\sum_\alpha \phi_{\alpha i}a_\alpha$. 
The coefficients $\phi_{\alpha i}$ describe the single-particle wave-functions $\phi_\alpha$ that correspond to the eigenvectors of the matrix
\begin{equation}
\label{Mmatr}
M_{ij} = \begin{cases}
          \omega_i & i=j,\\
          -t & i,j \; \mbox{ nearest neighbours,}
         \end{cases}
\end{equation}
with eigenvalue $\epsilon_{\alpha}$. They verify the orthogonality and completeness relations, i.e., 
$\sum_i \phi^*_{\alpha i} \phi_{\beta i}=\delta_{\alpha\beta}$ and $\sum_{\alpha} \phi^*_{\alpha i} \phi_{\alpha j} = \delta_{ij}$.
In terms of $a_\alpha, a_\beta^\dagger$, one obtains the diagonal Hamiltonian
\begin{equation}
\label{AndersonDiag}
\HH{A} = \sum_{\alpha=1}^{N_e} \epsilon_\alpha a^{\dagger}_\alpha  a_{\alpha}.
\end{equation}
The many-body eigenstates of \eqref{AndersonDiag} can be written as
\begin{equation}
\label{eigenmu}
\ket{\mu}=\ket{n^\mu_{1},n^\mu_{2},...,n^\mu_{N_e}} = \prod_{\alpha=1}^{N_e} (a_{\alpha}^\dag)^{n^\mu_{\alpha}} \ket{0}\;,
\end{equation}
where $n^\mu_{\alpha} \in \{ 0,1\}$ is the occupation number of the single-particle mode $\alpha$ in the many-body eigenstate $\ket{\mu}$.
Their energy is given as $E_\mu = \sum_\alpha n^\mu_{\alpha} \epsilon_{\alpha}$.
This is a huge simplification as compared to the original spin problem, since determining the many-body eigenstates of $\HH{A}$ only requires the numerical diagonalization of the matrix $M_{ij}$ in \eqref{Mmatr},
whose size grows 
linearly with the volume of the system, in contrast to the exponential growth of the Hilbert space of the spin model.

\begin{figure}
\includegraphics[width=0.49\textwidth]{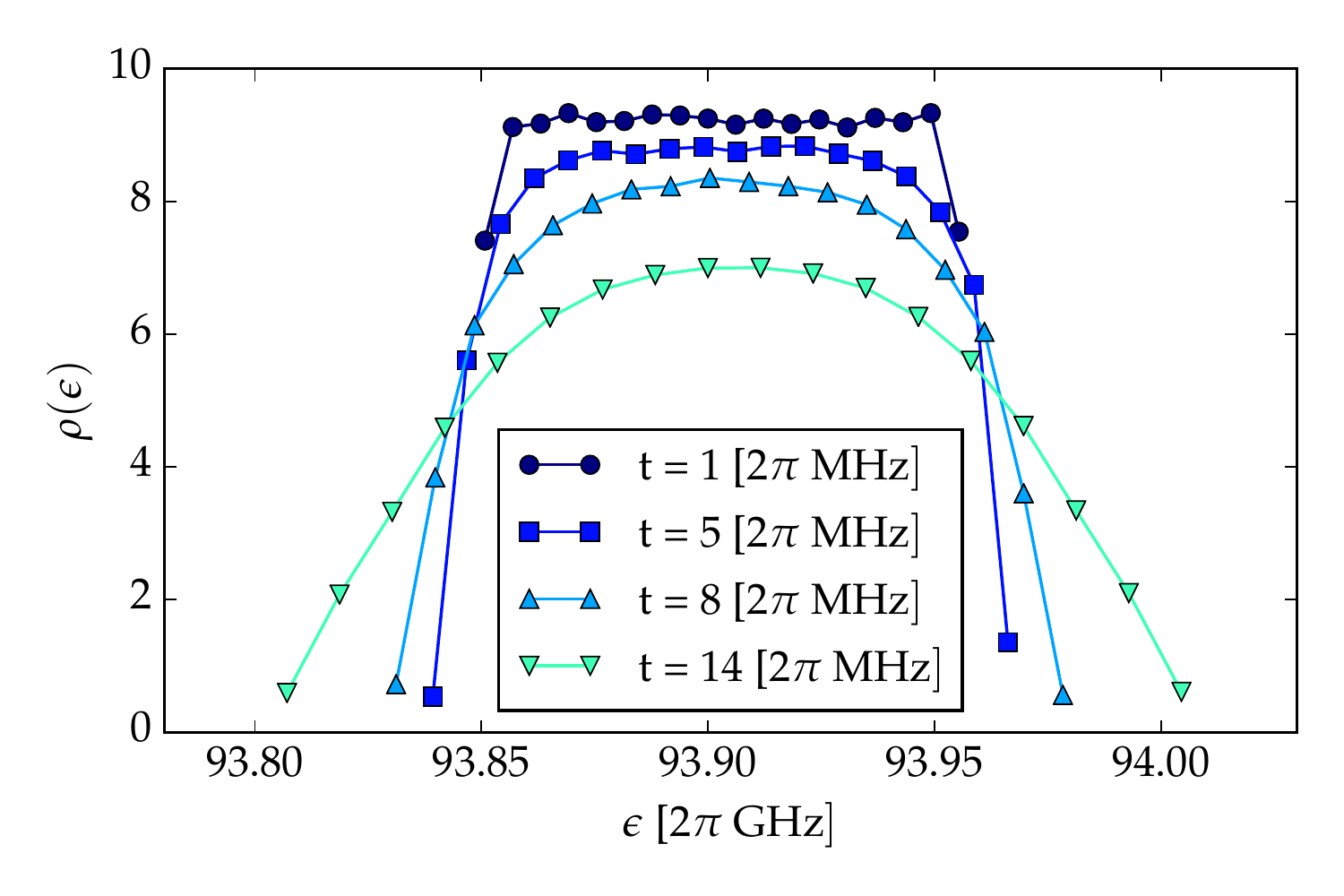} 
\caption{Density of states of the $3D$-Anderson model for a system of linear size of $L=8$ at constant disorder strength $w = 108 (2\pi\mbox{MHz})$ and for different values of the hopping parameter $t=1,5,8,14$ ($2\pi$MHz), corresponding to inverted and regular triangles, squares and dots respectively. }
\label{DOS}
\end{figure}

In our simulations, we consider a cubic lattice of linear size $L$, corresponding to $N_e = L^3$
spins in the original model. Then, the Anderson transition occurs at $t_c= w/16.5 \approx (6.6 \pm 0.2)$  ($2\pi$MHz). 
In Fig.~\ref{DOS}, the density of states (DOS) $\rho(\epsilon) = N_e^{-1}\sum_{\alpha} \delta(\epsilon - \epsilon_\alpha)$
is shown for different values of $t$: this is a self-averaging quantity which is not affected 
by the localization transition; for large $t$, its shape 
is broader and smoother than the box distribution obtained in the absence of hopping, i.e., in the atomic limit. 

\subsection{Relaxation and microwave dynamics in the free-fermion model}
 In the DNP protocol, the system is in contact with a thermal bath at a temperature $\beta^{-1}$
but driven to an out-of-equilibrium stationary state by microwave irradiation. 
Both bath and microwaves induce spin flips in the system. Each flip causes transitions between two eigenstates of the system, $\ket{n}$ and $\ket{m}$.
The general expressions for the corresponding transition rates were derived in [\onlinecite{DeLuca2015},\onlinecite{deluca2016}].
After the translation to fermions, a spin flip at a given site corresponds to the injection or emission of a fermion at that site.
This is a rather artificial process, however, since the drive injects or extracts fermions at a sharply defined energy. To realize this physically one would require a large fermionic reservoir with a very narrow energy filter. However, we remind the reader that we consider the fermionic model primarily as a theoretical toy model, with the main aim to study correlation functions analogous to those arising in the more realistic spin model.

Due to the factorization in \eqref{factorization}, the state of the nuclear spins, $\ket{A}$ is unchanged in these processes.
Using \eqref{eigenmu}, the matrix elements for spin raising or lowering operators translate into single particle matrix elements for fermionic creation or annihilation operators. The corresponding transition rates for processes due to bath coupling or  driving are then given by (see Appendix \ref{mathRates} for a brief derivation):
\begin{subequations}
\label{transRate}
\begin{align}
W^{\text {\tiny BATH}}_{\mu \to \nu}
&=
\sum_{\alpha} \frac{h(\Delta E_{\mu \nu})}{T_{1e}} \; \left(|\bra{\mu} a_\alpha^\dagger \ket \nu|^2 + |\bra{\mu} a_\alpha \ket \nu|^2\right)\;, 
\\
W^{\text {\tiny MW}}_{\mu \to \nu}
&=
\sum_{\alpha} \frac{T_{2e}\omega_1^2  \left(|\bra{\mu} a_\alpha^\dagger \ket \nu|^2 + |\bra{\mu} a_\alpha \ket \nu|^2\right)}
{1+T_{2e}^2\left(|\Delta E_{\mu\nu}| - \omw \right)^2}  \;
 \;.
 \end{align}
\end{subequations}
They govern the master equation describing the dynamics of occupations of many body eigenstates. 
Here, the function $h(\epsilon)=e^{\beta\epsilon}/(1+e^{\beta\epsilon})$ assures the detailed balance 
characteristic of a thermal equilibrium at temperature $\beta^{-1}$.  $T_{1e}$ and $T_{2e}$ 
are, respectively, the relaxation and coherence times for the system of electron spins, and $\Delta E_{\mu\nu}=E_\mu-E_\nu$.
Note that in order for these rates to be non-vanishing, the states $\ket{\mu}$
and $\ket{\nu}$ have to differ  in one and only one single mode occupation $n_\alpha$. Therefore,
we only need to know the single-particle energies $\epsilon_\alpha$
to obtain explicit expressions for the rates in \eqref{transRate}. 
As a consequence, in the absence of the hyperfine couplings, the occupation dynamics of different single-particle modes $\alpha$ 
 decouple. Denoting by $\hat n_\alpha = a^\dag_\alpha a_\alpha$  the occupation number operator for the mode $\alpha$, 
we can obtain its stationary expectation value as $\langle \hat n_\alpha\rangle_{t\to \infty} = (1 + P_B(\epsilon_\alpha))/2$, with the bias
\begin{align}
\label{pbloch}
P_B(\epsilon)=&\frac{\left[1+T_{2e}^2 (\epsilon-\omw)^2 \right]\times \tanh 
(\beta  \epsilon/2)}{1+T_{2e}^2 (\epsilon-\omw)^2+2 \omega_1^2 T_{1e} T_{2e}} \;.
\end{align}
This expression coincides with that for the stationary polarization of a single-spin
coupled to a bath and driven by microwave irradiation, known as the Bloch equation. 
From these biases one immediately obtains the expression for the stationary occupation 
probability of any eigenstate $\ket{\mu}$ as
\begin{equation}
\label{pmustat}
p_\mu^\text{\tiny stat} = \frac{1}{2^{N_e}}\prod_{\alpha=1}^{N_e} [1 - (-1)^{n_{\alpha}^\mu} P_B(\epsilon_\alpha)].
\end{equation}

\subsection{Rates involving nuclear transitions}
The hyperfine interactions $\HH{e--n}$ in \eqref{hHF}
induces two types of transitions (see Appendix~\ref{hfMath} for the details): 
\paragraph A three-spin transition labelled \emph{ISS} which is responsible for the \emph{cross-effect} mechanism of hyperpolarization. 
This transition involves the spin-flip of a nuclear spin
together with the \emph{flip-flop} of two electron spins. In our free-fermion model, 
the flip-flop transition can connect eigenstates $\ket\mu$ and $\ket\nu$ 
only if they differ in the occupation number of two single-particle modes, as in the example:
\begin{equation}
\label{ISSexample}
\ket{A} \times
\ket{0,\underbrace{\bf 1}_{\alpha},1,\underbrace{\bf 0}_\beta} \; \rightleftharpoons\; 
\ket{B} \times
\ket{0,\underbrace{\bf 0}_\alpha,1,\underbrace{\bf 1}_\beta}\;,
\end{equation}
where $\ket A$ and $\ket B$ are two eigenstates of $\HH{n}$, whose features will be specified later on.
The rate of the process in \eqref{ISSexample} involving the spin-flip 
of the nucleus $j$
is
\begin{equation}
\label{Wflipflop}
W^{\text{\tiny ISS}}=
\frac{T_{2n}A_{j}^2 |\phi_{\alpha,j}|^2 |\phi_{\beta,j}|^2 |\braket{A|\hat I_j^x|B}|^2}{1+T_{2n}^2\left(|\epsilon_\alpha-\epsilon_\beta| - \omega_n\right)^2}  
\;,
\end{equation}
where $T_{2n}$ denotes the coherence time of a nuclear spin. 
This channel becomes very efficient when the resonance condition $|\epsilon_\alpha-\epsilon_\beta| \sim \omega_n$ is matched.
\paragraph A \emph{polarization-loss} (PL) transition where a nuclear spin flip leaves 
the many-body fermionic state unchanged:
\hspace{0.5cm} 
\begin{equation}
\label{LEAKexample}
\ket{A}\times \ket{\mu} \; \rightleftharpoons\; \ket{B} \times\ket{\mu}
\end{equation}
for any fermionic eigenstate $\ket{\mu}$. 
The rate of the process in \eqref{LEAKexample} induced by the operator $\hat{I}_j^x$  writes:
\begin{equation}
\label{WPL}
W^{\text{\tiny PL}}=
\frac{T_{2n}A_{j}^2 |\braket{A|\hat I_j^x | B}|^2}{1+T_{2n}^2\omega_n^2} \left [ \sum_\alpha |\phi_{\alpha,j}|^2 \bigl( n_\alpha^\mu -\frac{1}{2} \bigr) \right ]^2
 \;.
\end{equation}
Note that this term is always off-resonant and therefore its effect is rather weak as compared to the other transitions. 
In particular, dissipative processes 
that decrease nuclear polarization
are governed by the contact with the bath, which leads to the leakage rate
\begin{equation}
\label{Wleakage}
W^{\text {\tiny LEAK}}_{A \to B}
=
\sum_{p=1}^{N_n} \frac{h(\Delta E_{A B})}{T_{1n}} \; |\bra{A} \hat I_p^x \ket B|^2 \;,
\end{equation}
with $h(x)$ defined below Eqs.~\eqref{transRate} and 
$T_{1n}$ being the relaxation time of a nuclear spin due to contact with the bath. 
These processes are typically more relevant than $W^{\text{\tiny PL}}$, as they affect  
all nuclei and not only those coupled to electrons.

\section{Inducing nuclear hyperpolarization\label{Sec:Results}}
Equations \eqref{transRate}, \eqref{Wflipflop}, \eqref{WPL} and \eqref{Wleakage} define the rate 
equations governing the occupation of the many-body eigenstates of the free fermion model which mimicks the spin system. 
The stationary state is given by the
occupation probabilities $\pnstat$, obtained as the 
eigenvector associated with the vanishing eigenvalue of the \emph{transition matrix}
\begin{equation}
T_{n,n'} = W^{\text{\tiny tot}}_{n'\to n} - \delta_{n,n'} \sum_{m} W_{n \to m}^{\text{\tiny tot}}\;,
\end{equation}
where $W^{\text{\tiny tot}}_{n'\to n}$ is the sum of all rates for processes taking $\ket{n'}$ to $\ket{n}$. 
The size of this matrix is huge, as it scales as $2^{N_e+N_n}$. 
However, the simple structure of many-body eigenstates based on the factorization~\eqref{factorization}
and on the free-fermion approximation allows in principle 
the use of a Markov-chain Monte Carlo method with a complexity that grows only linearly with the system size. 
However, we will not pursue this strategy here, but leave it for future studies.

In the following, we are interested in a situation as it arises in DNP protocols with trityl radicals, for which hyperfine couplings are weak. In such cases, the stationary state of the electron  spins was empirically found to be independent of the concentration of nuclear spins.\cite{ColomboSerra2014} 
This suggests that the occupation probability $p_{A,\mu}$ 
of having the nuclear spin system in the state $\ket A$ and the fermions in the many-body eigenstate $\ket{\mu}$ can be factorized, 
$p_{A, \mu} = p_{A} \times p_{\mu}$. Then, 
in the limit of vanishing hyperfine interaction, 
we can assume  the electron system to permanently remain  in the stationary state, so that 
we can replace $p_{\mu} \to p_\mu^\text{\tiny stat} $ in \eqref{pmustat}.
This allows us to trace out the electron spin degrees of freedom and obtain a rate equation
 for the nuclear spins only.

\subsection{Stationary value of the nuclear polarization} 
In the presence of efficient nuclear dipolar interactions 
(as usually given in experiments), 
we assume that statistical properties of nuclear many-body eigenstates are completely characterized by global conserved quantities. 
In the present case this is only the total  energy of the system of nuclear spins, $\HH{n}$. 
Therefore, it suffices to describe the evolution of the total energy of the nuclear system.
For simplicity, we take the same hyperfine coupling $A_j = A_0$ in \eqref{hHF}
for all ``core'' nuclear spins that are coupled to the electrons, and set $A_p=0$ for all other nuclei.
Given that $\omega_n$ is much larger than the typical strength of dipolar coupling between nuclear spins,
we assume that dipolar interactions are strong enough to render them  ergodic, while contributing only a negligible broadening to the Zeeman gap $\omega_n$.
In this limit, the energy of a nuclear eigenstate $\ket{A}$ only depends on the number of up/down spins:
$E_A = (N_+^{(A)} - N_-^{(A)})\omega_n/2$. The total rate of transitions that lower the nuclear energy by $\omega_n$ 
can be obtained as $\Omega_{\rm emit} =\sum_p \Omega_p(\omega_n)$, where the function $\Omega_j(\omega)$ is derived in Appendix \ref{hfMath} and reads
\begin{align}
\label{OmegaFinal}
 \Omega_p(\omega) = &\frac{A_{p}^2 T_{2n}}{4}\sum_{\alpha,\beta} |\phi_{\alpha,p}|^2 |\phi_{\beta,p}|^2\times 
 \Bigg[\frac{P_B(\epsilon_\alpha) P_B(\epsilon_\beta)}{1+T_{2n}^2\omega^2} +\nonumber\\
&+ \frac{ (1+P_B(\epsilon_\alpha)) (1-P_B(\epsilon_\beta))}{1+T_{2n}^2(\omega+\epsilon_\beta-\epsilon_\alpha)^2}\Bigg] + \frac{h(\omega)}{T_{1n}} \;.
\end{align}
Likewise the rate of transition for an absorption process (that rises the energy of $\omega_n$) is $\Omega_{\rm abs}= \sum_p \Omega_p(-\omega_n)$.
Note that with this expression we also take into account the contribution of nuclear spins
not coupled to the electrons, for which only the leakage part is present as $A_p = 0$.


Since the nuclear spins establish an equilibrium with temperature $\beta_N^{-1}$ among themselves, the balance of energy flows in the steady state requires that   $\exp(-\beta_N\; \omega_n)= \Omega_{\rm abs}/\Omega_{\rm emit}$, from which one deduces the 
total magnetization 
\begin{equation}
\label{pnETH}
P_n= \tanh\left (\frac{\beta_N \;\omega_n}{2}\right ) = \frac{\Omega_{\rm emit} - \Omega_{\rm abs}}{\Omega_{\rm emit} + \Omega_{\rm abs}}\;.
\end{equation}
An alternative derivation of this result including a non-trivial broadening due to the interactions
is presented in appendix~\ref{ETHMath} using Srednicki's ETH formula, \cite{srednicki1999approach} for 
the matrix elements of local operators in an ergodic system. 

\begin{figure}
\includegraphics[width = 0.45\textwidth]{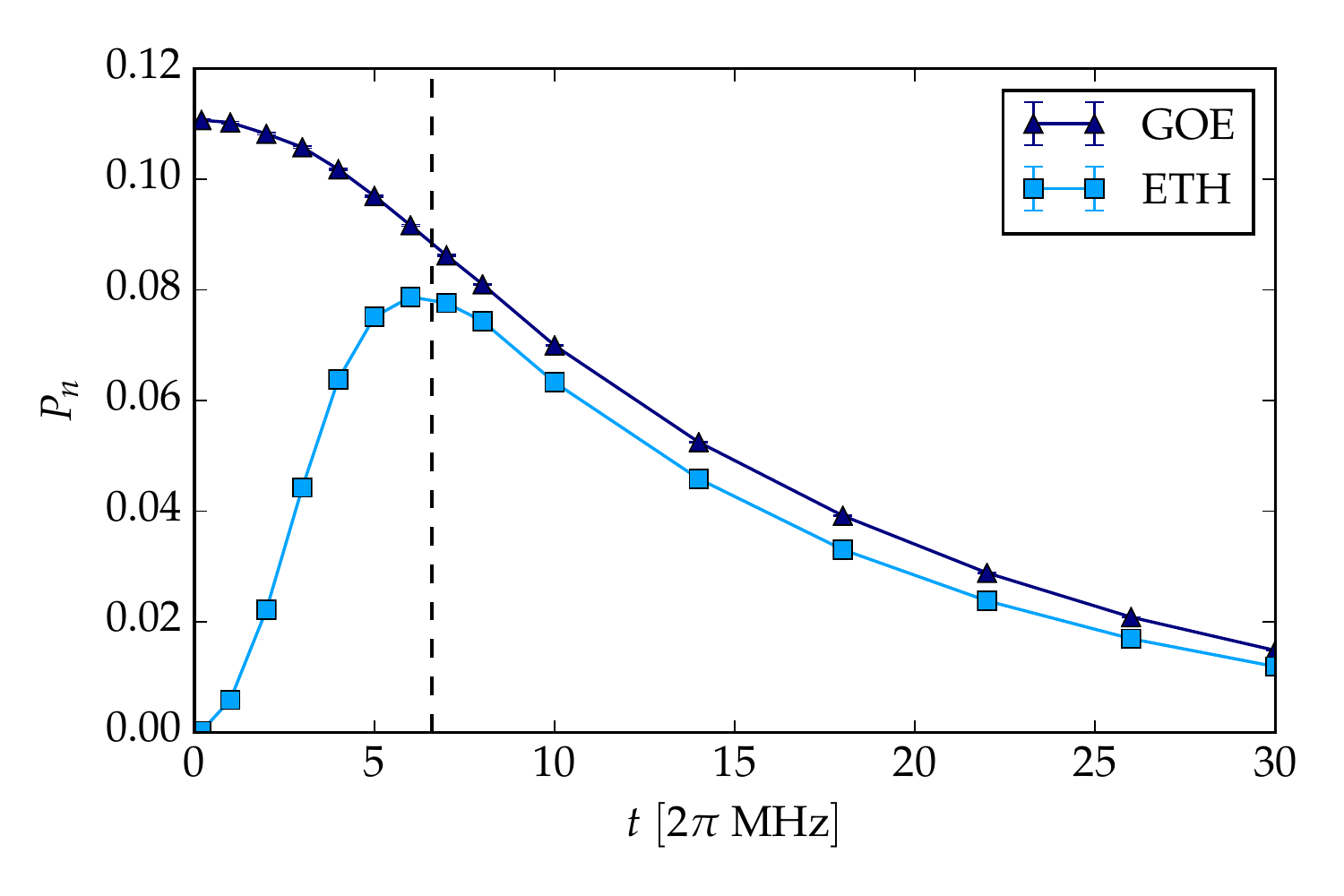} 
\caption{(Color online) Average nuclear polarization from Eq.~\eqref{pnETH} with the overlap function  from Eq.~\eqref{OvFunc} (light blue squares), contrasted with the result obtained by assuming GOE-like electron wavefunctions (cf Eq.~\eqref{KGOE}) (blue triangles). We have a total of $N_n$ nuclei, of which $N_e$ are coupled to a fermion site in the $3D$ Anderson Model. The vertical line identifies the known Anderson transition. The linear size of the system was $L=18$. Data are averaged over $60$ realizations of the disorder and error bars indicate one standard deviation. 
}
\label{PnVsGOE}
\end{figure}

\begin{figure*}
\includegraphics[width = 0.45\textwidth]{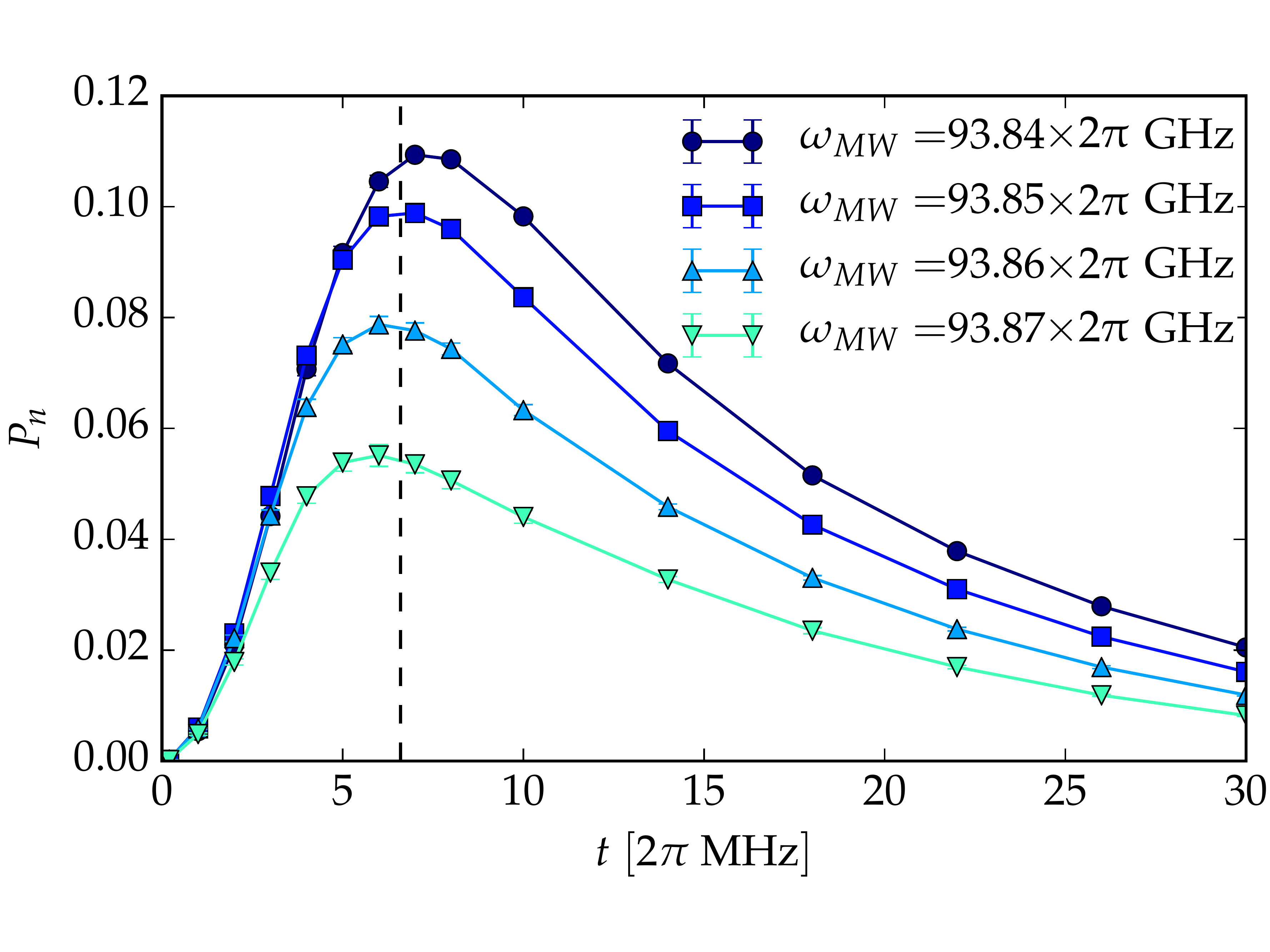} 
\includegraphics[width = 0.45\textwidth]{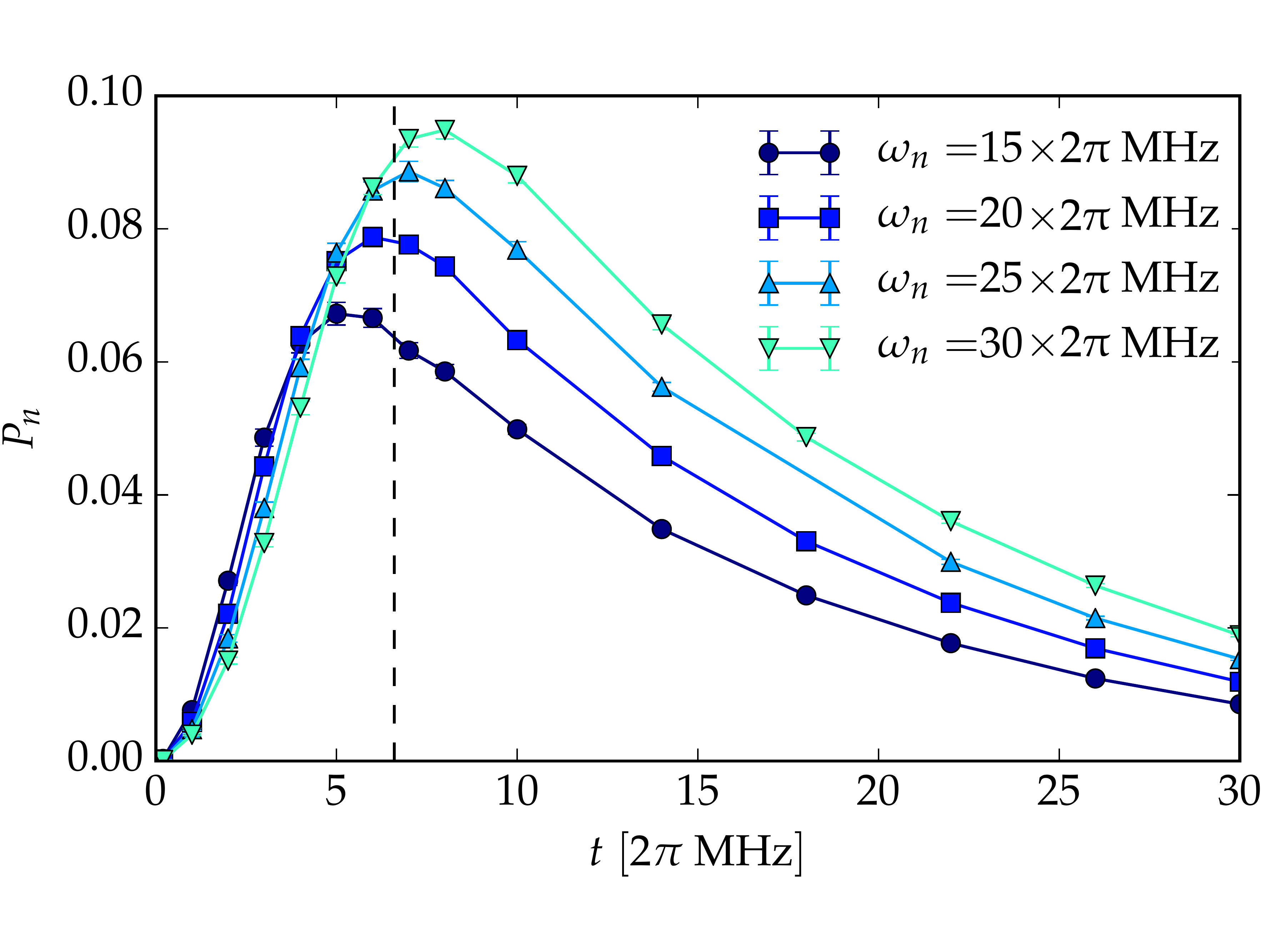} 
\caption{(Color online)  
Average polarization obtained from Eq.~\eqref{pnETH} as a function of the hopping parameter $t$. The maximal polarization is always obtained close to the localization transition, $t\approx t_c$. 
(left) Different values of the microwave frequency $\omw$ (shifting in steps of $10\times2\pi$~MHz towards the center of the disorder distribution at $\omega_e=93.9\times 2\pi$~GHz), with fixed $\omega_n = 20\times 2\pi$~MHz. 
(right) Different values of $\omega_n$, corresponding to different nuclear spin species, fixing $\omw=93.86\times 2\pi$~GHz.
For both plots, the vertical line identifies the known Anderson transition in the undriven system, assuming a fixed width of the box distribution of disorder, ${w}= 108 (2 \pi\mbox{MHz})$. 
The linear size of the system was $L=18$. Each data point was obtained by averaging over $60$ realizations of disorder. The error bars
correspond to one standard deviation. 
\label{PNMWWN}}
\end{figure*}

\begin{table}
\begin{center}
\begin{tabular}{|c|c|c|c|}
\multicolumn{4}{c}{Parameters of electron spins}\\
\hline
  $T_{1e}$ (s) &  $T_{2e}$ (s) &  $\omw$ ($2\pi$GHz)  & $\omega_1$ ($2\pi$kHz) \\
  \hline
  $1$ & $10^{-6}$ & $93.86$ & $25$      \\
 \hline 
 \multicolumn{4}{c}{ }\\
 \multicolumn{4}{c}{Parameters of nuclear spins}\\
 \hline 
  $T_{1n}$ (s) &  $T_{2n}$ (s) & $\omega_n$ ($2\pi$MHz)  & $A_0$ ($2\pi$kHz)  \\
  \hline 
  $10^4$ & $10^{-5}$ & $20$ & $4.47$    \\
 \hline  
\end{tabular}\\
\caption{Microscopic parameters pertaining to a trytil system in a magnetic field of $B=3.3$ T ($\omega_e=93.9$ $2\pi$GHz) 
in contact with a thermal bath at temperature $\beta^{-1}=1.2$ K and in the presence of microwave irradiation with amplitude $\omega_1$.
The hyperfine coupling is assumed small enough so that the transitions involving nuclei 
are slower than the relaxation time $T_{1e}$ of electron spins: $A_0^2/(\omega_n^2 T_{2n}) \lesssim T_{1e}^{-1}$. 
This justifies the approximation that the electron spins remain in their steady state $p_\mu \simeq p_\mu^\text{\tiny stat} $
at all times. Stationary nuclear polarizations only depend on $A_0^2 T_{1n}$. 
}
\label{tableParam}
\end{center}
\end{table}

\begin{figure}
\includegraphics[width = 0.45\textwidth]{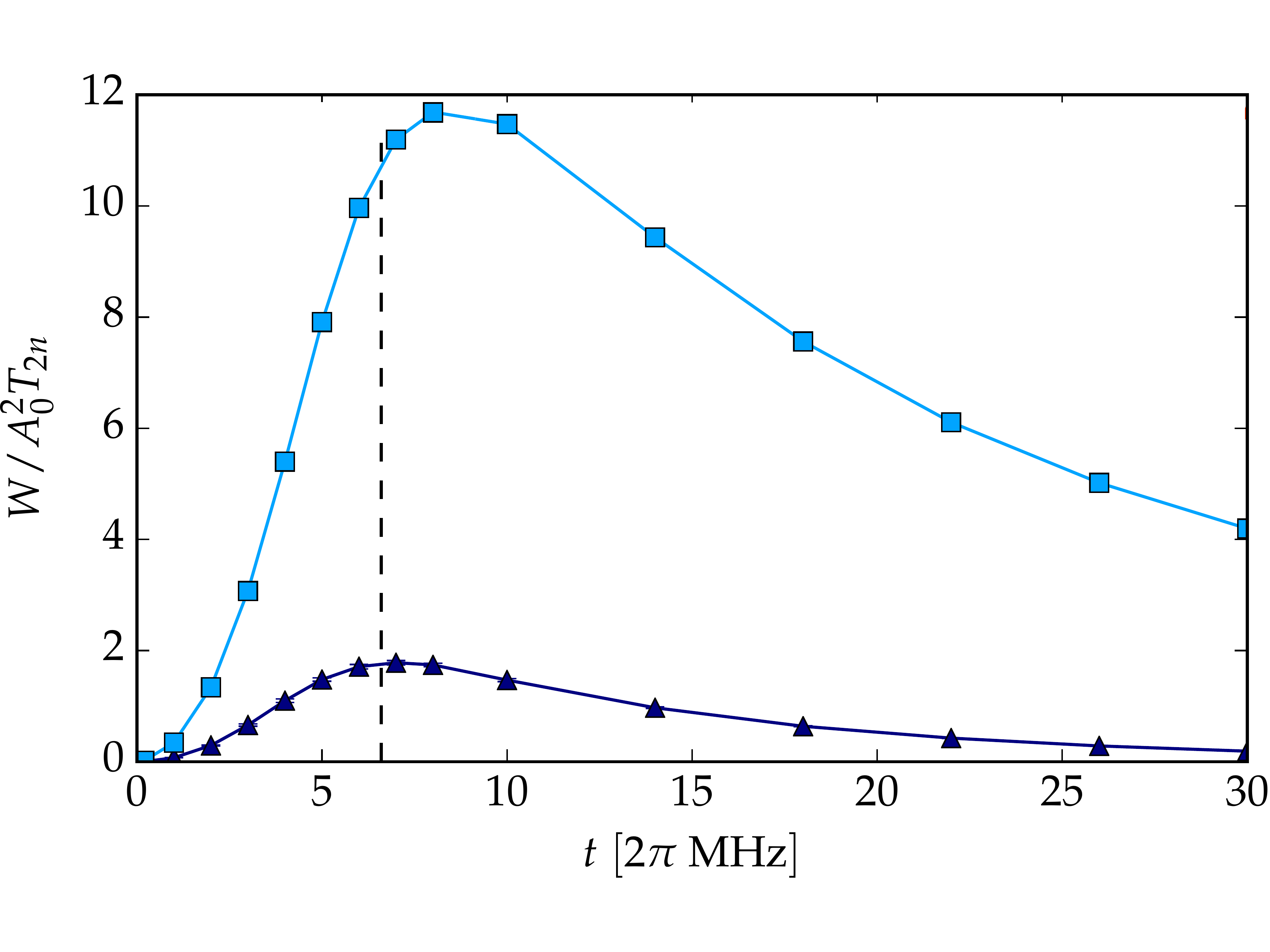} 
\caption{(Color online) Comparison between the different (normalized) rates appearing in \eqref{polnucerg}, as a function of $t$: 
The difference of emission and absorption rates in the numerator (dark blue triangles) and
their sum (the first term in the denominator), corresponding to the ISS processes (light blue squares)
}
\label{NUMDENCOMPI2}
\end{figure}

An important aspect is that the microscopic details of the electron system
determine the stationary nuclear polarization \eqref{pnETH} 
only via the correlation function
\begin{equation}
\label{OvFunc}
\mathcal{K}(\epsilon, \epsilon') = \frac{1}{N_e}\sum_{\alpha, \beta, j} \overline{|\phi_{\alpha, j}|^2 |\phi_{\beta, j}|^2 \delta(\epsilon_\alpha - \epsilon) \delta(\epsilon_\beta - \epsilon')} \; ,
\end{equation}
which measures the overlap between eigenvectors at energies $\epsilon$ and $\epsilon'$. This correlator was discussed in 
detail in [\onlinecite{cuevas2007two}] in the context of Anderson localization, where it characterizes multifractality and the strong correlation in the spatial support of wavefunctions with close energies. 
 
Had we worked within the original spin model only, we would have found that the relevant correlator appearing in (19) is just the local spin-spin correlator $S_p^z(\omega)S_p^z(-\omega)$. Its behavior controls the nuclear spin polarization.  Similar correlators of local observables have also been proposed as
\cite{serbyn2017} as a possible way to characterize multifractality in 
in many-body systems across the MBL crossover; however, their quantitative analysis is problematic in general, so 
we analyze instead the related corrrelator $n_p(\omega)n_p(-\omega)$ in the solvable single particle fermionic model, that we believe to capture key qualitative ingredients of the (single-particle) delocalization transition. 

Using the above expression 
in \eqref{pnETH}, we obtain
\begin{equation}
\label{polnucerg}
P_n=\frac{\frac{N_e}{4}\int d\epsilon d\epsilon' \mathcal{K}(\epsilon, \epsilon')  \frac{P_B(\epsilon)- P_B(\epsilon')}
{1 + T_{2n}^2 (\epsilon- \epsilon' + \omega_n)^2}
}{
\frac{N_e}{4} \int d\epsilon d\epsilon' \mathcal{K}(\epsilon, \epsilon') 
\frac{1 - P_B(\epsilon')P_B(\epsilon)}{1 + T_{2n}^2 (\epsilon- \epsilon' + \omega_n)^2
} + \frac{2 N_n}{A_0^2 T_{1n} T_{2n}}
} \;,
\end{equation}
where we neglected the contribution from the polarization loss rate in \eqref{WPL},
since it is subleading with respect to the leakage term of Eq.~\eqref{Wleakage}. Indeed,
while the polarization loss only affects $N_e$ core nuclei, the leakage affects all  $N_n$ nuclei: Under
standard DNP conditions $N_n / N_e \simeq 10^3$ a value that we will use in the following. 

In Fig.~\ref{PnVsGOE}, we show the behavior of the nuclear polarization 
in \eqref{pnETH} as a function of the hopping parameter $t$. 
To elucidate the role played by localization, we compare with the polarization one obtains by assuming electronic single-particle eigenfunctions that are perfectly ergodic in the whole volume. To do so, we keep the original values of $\epsilon_\alpha$'s 
but replace the wavefunctions $\phi_{\alpha, i}$ obtained from the diagonalization of $M$ in \eqref{Mmatr}, with those obtained from the diagonalization of a matrix in the Gaussian orthogonal ensemble (GOE). In this way, the DOS remains unchanged, while the eigenstates are perfectly delocalized, with $|\phi_{\alpha i}|^2\approx1/N_e$.
This leads to a trivial correlator:
\begin{equation}
\label{KGOE}
\mathcal{K}^{\tinytext{GOE}}(\epsilon, \epsilon') = \rho(\epsilon) \rho(\epsilon') \;.
\end{equation}
Introducing this approximation in \eqref{polnucerg},
we obtain a monotonous behavior for the polarization, which we will rationalize below. 

As one should expect the two curves agree rather well the deeply delocalized regime, $t \gg t_c$. However, at small values of $t$,
leakage is responsible for a strong suppression and the nuclear polarization vanishes as $O(t^2)$. 
As we will explain in the next section, the nuclear polarization is suppressed both at small values of $t\ll t_c$ and for $t \gg t_c$. As a consequence,  the nuclear polarization 
reaches a maximal value at an intermediate value $\tmax$ of the hopping. More precisely, we will argue that in the proximity of the Anderson transition the polarization is particularly  enhanced and thus the maximum occurs close to the transition $\tmax \simeq t_c$.
In Fig.~\ref{PNMWWN}, this general behavior is confirmed for a broad range of microwave frequencies $\omw$ and nuclear Zeeman gaps $\omega_n$ (corresponding to different nuclear species).

\section{Behavior of the nuclear polarization \label{Sec:Discussion}}
\subsection{Nuclear polarization away from criticality}
\begin{figure}
\includegraphics[width = 0.45\textwidth]{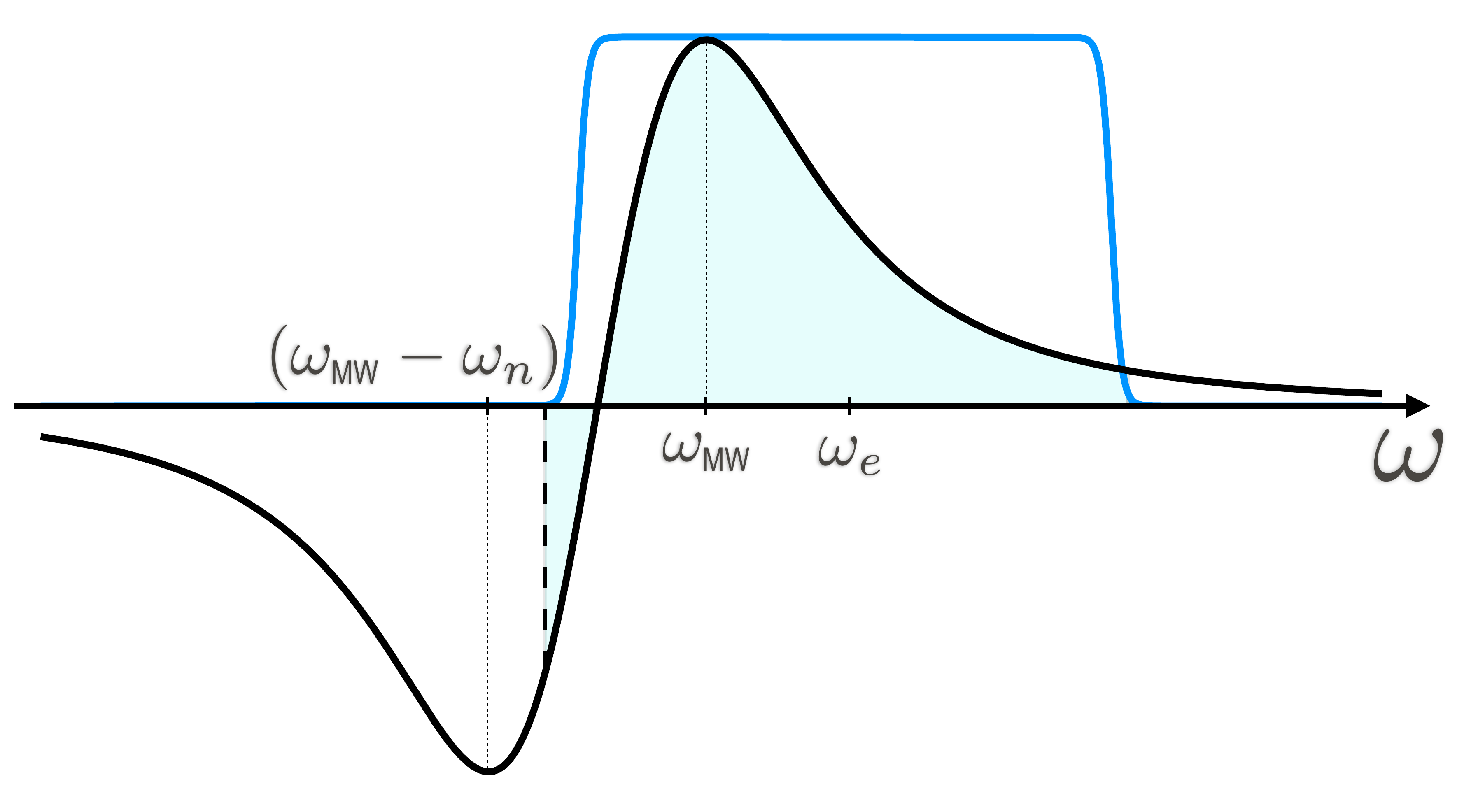} 
\caption{(Color online) 
Sketch of the factors in the integrand of \eqref{polnucergapprox}: $P_B(\epsilon) - P_B(\epsilon + \omega_n)$ (black) and
the product of two DOS $\rho(\epsilon)\rho(\epsilon+\omega_n)$ (blue) (The normalization is adjusted to produce a clear plot).
The stationary polarization in the GOE approximation $P_n^{\tinytext{GOE}}$ is obtained
as the integral over the domain where the product of the two DOS does not vanish. Upon increasing $t$, the DOS broadens and the integration window becomes larger. Accordingly 
$P_n^{\tinytext{GOE}}$ decreases, both because the total integrand gets closer to being odd and because of the decreasing magnitude of the DOS $\rho(\epsilon)$ in the bulk.
}
\label{DOSPBLOCH}
\end{figure}

To gain further understanding, we show in Fig.~\ref{NUMDENCOMPI2}, the interplay of the different rates entering the numerator and denominator of \eqref{polnucerg}. Note that in our regime of parameters, the denominator is always dominated 
by the leakage term. 
This suggests that the behavior of the polarization can be understood qualitatively by
focusing on the numerator of \eqref{polnucerg}. There, the contribution of the ISS process
is strongly peaked at the resonant condition $\epsilon - \epsilon' \simeq \omega_n$, which leads to
the following approximate formula for the polarization
\begin{equation}
\label{polnucergapprox}
P_n\simeq \frac{\pi A_0^2 T_{1n} N_e}{8 N_n}
\int d\epsilon \, \mathcal{K}(\epsilon, \epsilon + \omega_n)  (P_B(\epsilon)- P_B(\epsilon + \omega_n))
 \;.
\end{equation}
In the delocalized region, we can use the GOE approximation, i.e. the form \eqref{KGOE} of the correlator. The dependence  of the nuclear polarization on $t$ then derives merely from the broadening of the DOS induced by the hopping. In Fig.~\ref{DOSPBLOCH},
we show that the nuclear polarization obtained from \eqref{polnucergapprox}, depends on the shift between the center of the irradiation frequency $\omw$ and the
center of the DOS, $\omega_e$. Upon increasing $t$ beyond the disorder level $w$, the density of electronic states is broadened. 
This has two effects (see Fig.~\ref{DOS}): 
On the one hand, the total integrand becomes increasingly odd around $\omega_{MW}-\omega_n$, which tends to suppress the integral. On the other hand 
the density of states itself decreases as $\rho(\epsilon) \simeq O(t^{-1})$.
This large-$t$ depletion is independent of localization phenomena and is at play mostly as long as $t\gtrsim w$, 
that is, well into the ergodic phase. A similar phenomenon was already discussed in~[\onlinecite{deluca2016}].

The localized regime and critical regimes are more interesting. Indeed, the eigenfunction correlator $\mathcal{K}(\epsilon, \epsilon')$ is affected by the onset of localization. 
To explain qualitatively its behavior, we remark that in the presence of a flat disorder distribution, as considered here, 
it essentially depends on the energy difference only, as
\begin{equation}
\label{Kdiff}
\mathcal{K}(\epsilon, \epsilon+ \omega) \simeq \rho(0)^{-1} K(\omega) \;, \quad K(\omega) = \int d\epsilon \; \mathcal{K}(\epsilon, \epsilon + \omega)\;.
\end{equation}
where $\rho(0)^{-1} \simeq \max(t, w)$ is the bandwidth of the DOS.
As one can check explicitly from \eqref{OvFunc}, it must satisfy the sum rule
\begin{equation}\label{sumrule}
 \int d\omega\,  K(\omega) = 1\;. 
\end{equation}
To proceed further, it is convenient to define the \textit{inverse participation ratio} (IPR) of single particle wavefunctions as
\begin{equation}
\label{IPR2}
 I_2 = \frac{1}{N_e}\sum_{\alpha,j} \overline{|\phi_{\alpha,j}|^{4}}\;.
\end{equation}
Once  localization has set in, 
single-particle eigenstates have a support
over a finite number of lattice sites whose number can be measured by $I_2^{-1}$. Then,
the overlap function $K(\omega)$ becomes singular at small $\omega$, i.e.
\begin{equation}
K(\omega) = \delta(\omega) I_2 + \mbox{\textrm{non-singular contributions}}.
\end{equation}
Since at small $t$, the sum rule \eqref{sumrule} is almost saturated by $I_2$ at $\omega = 0$,
$K(\omega)$ for $\omega \simeq \omega_n$ is strongly suppressed.  This  expresses the simple fact, that at small hopping amplitude wavefunctions with an energy difference exceeding $t$ are very unlikely to overlap in space.
As a consequence, the number of ISS processes is strongly reduced once the transition is crossed. The total polarization is found to scale as $O(t^2)$, as we will explain below.  
In the next subsection, we discuss the features of $K(\omega)$ that are reflected in the stationary nuclear polarization, especially close to the transition.

\subsection{Enhancement of nuclear polarization from multifractality at the localization transition}

\begin{figure}
\includegraphics[width = 0.45\textwidth]{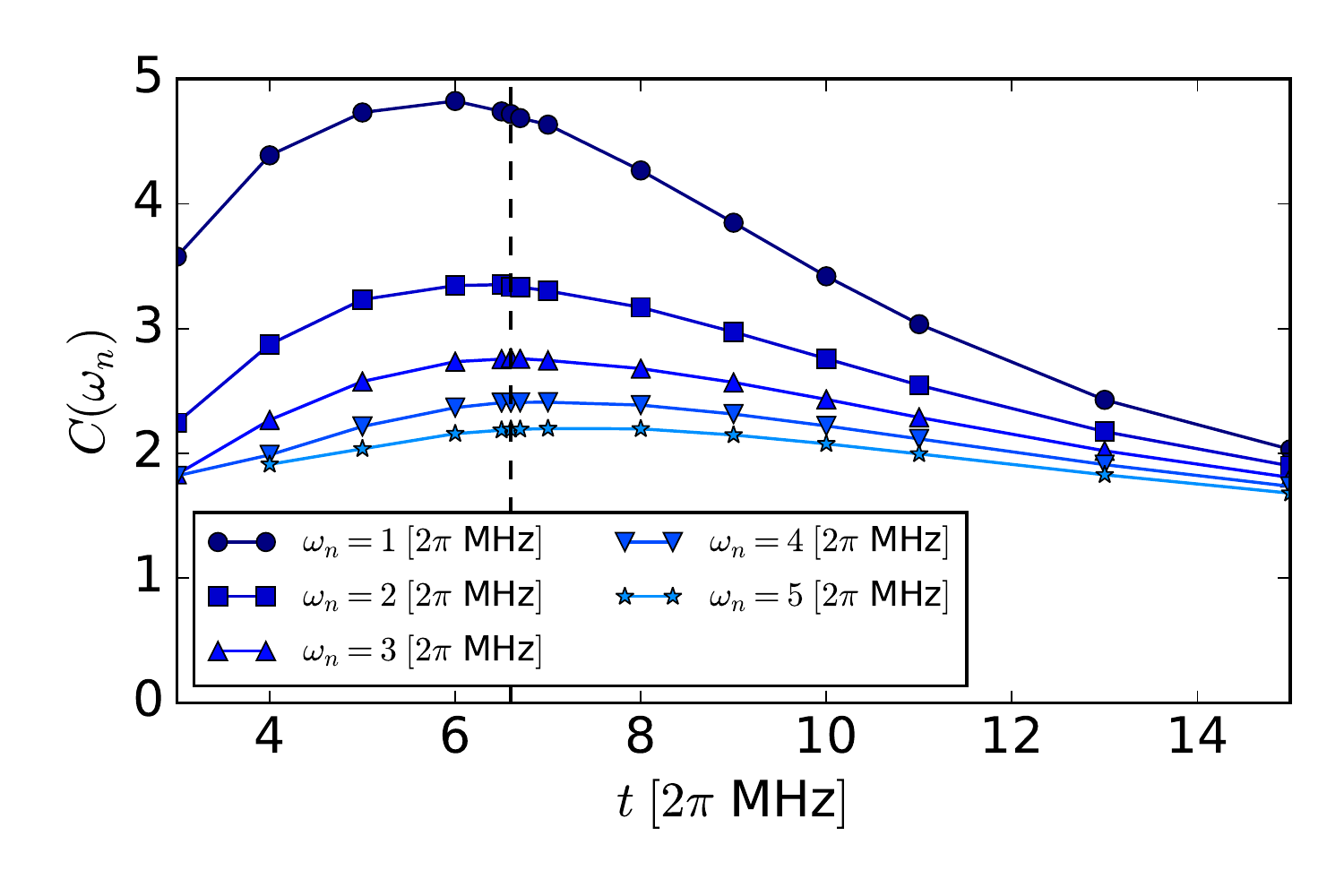} 
\caption{(Color online) Eigenfunction correlation function $C(\omega_n)$ defined in Eq.~\eqref{Cdef} as a function of the hopping strength $t$ for small values of the nuclear Zeeman energy $\omega_n=1,2,3,4,5$ ($2\pi$~MHz). 
This overlap function controls the nuclear polarization in the localized regime. It peaks very close to the Anderson transition.
The linear size of the system was $L=20$ and each point of the hopping was obtained by averaging over $10$ realizations of the Anderson Hamiltonian in Eq.~\eqref{hAnd}. }
\label{OVERLAP}
\end{figure}

In the previous section, we observed that, in general, 
the stationary nuclear polarization is expected to decay
both at $t\ll t_c$ and $t\gg t_c$, thus reaching a maximum at an intermediate value $\tmax$. We now argue
that, for a broad range of frequencies $\omega_n$, this value $\tmax$ is close to $t_c$, meaning that a maximal nuclear polarization is achieved by tuning the electronic system close to the Anderson transition. We had reported a similar effect for interacting spins in previous work on very small systems.~\cite{deluca2016} Here, however, we can reach a much deeper understanding of the role played by the localization transition.

For simplicity, we assume that $T_{1n}$ and $T_{2n}$ are such that i) the leakage in \eqref{Wleakage} dominates the denominator of \eqref{polnucerg} and ii) the integrand of \eqref{polnucerg} is dominated by the contributions at $\epsilon - \epsilon' \simeq \omega_n$. Then, we can write
\begin{equation}
\label{POverlapLink}
 P_n \simeq C(\omega_n) \times P_n^{\tinytext{GOE}} 
\end{equation}
where we rescaled the overlap function by the DOS, setting
\begin{equation}
\label{Cdef}
 C(\omega) = \frac{K(\omega)}{\int d\epsilon \rho(\epsilon) \rho(\epsilon + \omega) } \;.
\end{equation}
From \eqref{POverlapLink}, it is clear that 
$P_n^{\tinytext{GOE}} $ is responsible for the decay at large $t/w$, through the broadening of the DOS.
In contrast, the function $C(\omega_n)$ controls the suppression at small $t$.
Indeed, at small values of $t/\omega_n, \, t/w \ll 1$, we can use perturbation theory for the wavefunctions. The overlap of non-resonant wavefunctions centered on nearest neighbors then scales as $t$, the square of which dominates $K(\omega)$. This immediately  yields
 the quadratic behavior $C(\omega_n) \simeq t^2/\omega_n^2$.

To understand qualitatively the behaviour of  $C(\omega)$ close to the transition, $t \simeq t_c$, we remark that, in the localized phase $ t < t_c$, the eigenfunctions $\phi_{\alpha j}$ decay exponentially on a  scale $\xi_{\mbox{\tiny loc}}(t)$, known as the localization length. Similarly, in the delocalized phase, one can introduce a correlation length characterizing the eigenfunctions, $\xi_{\mbox{\tiny del}}(t)$. Approaching the critical point $t \to t_c$, from the two sides, these lengths diverge. From the two sides, we define a length scale
\begin{equation}
\xi(t) \equiv \begin{cases} 
\xi_{\mbox{\tiny loc}}(t) & t < t_c  \;, \\
\xi_{\mbox{\tiny del}}(t) & t > t_c 
\end{cases}
\propto |t - t_c|^{-\nu},
\end{equation}
where the critical exponent $\nu$ has been introduced. Inside the \textit{critical volume} $V_\xi \equiv \xi(t)^d$ ($d=3$ being the spatial dimension) the statistics of eigenfunctions is governed by multifractality. It is reflected in the inverse participation ratio\cite{evers2008anderson}
\begin{equation}
\label{I2fract}
I_2 \simeq \begin{cases}
 \xi(t)^{-d_2} & t   \neq t_c, \\
 N_e^{-d_2/d} & t = t_c.
 \end{cases}
\end{equation}
For the Anderson model in $d=3$, we have $d_2 \approx  1.44$ and $\nu \sim 1.571$~\cite{slevin2014, cuevas2007two}.
Multifractality also affects the frequency dependence of the overlap function, which is governed by the same fractal dimension $d_2$ as $I_2$\cite{mirlin2000statistics}, via an  exponent $\mu = 1 - d_2/d \approx 0.52$. Slightly on the delocalized side, $t>t_c$, the overlap function behaves as
\begin{equation}
\label{KravtsovScenario}
C(\omega) \simeq \begin{cases}  \left(\frac{t}{\Delta_\xi}\right)^{\mu} \sim \xi^{d-d_2} \;,& \omega \ll \Delta_\xi,  \\
				   \left(\frac{t}{\omega}\right)^{\mu} \;, & 
				\Delta_\xi \ll \omega \ll 2t. 
                  \end{cases} \; 
\end{equation}
where $\Delta_\xi  = (\xi(t)^d \rho(0))^{-1}$ denotes the mean level spacing in the correlation volume. An analogous critical behavior holds for $t \lesssim t_c$, where $\Delta_\xi$ simply has to replaced by the mean-level spacing in the localization volume. 
The power-law decay  at intermediate values of $\omega$ reflects critical behavior, known as Chalker-Daniell scaling~\cite{chaklerdaniell, chalker1990scaling}.

 A thorough numerical study of the overlap function in the 3D Anderson model was performed 
in \citei{cuevas2007two} confirming the validity of \eqref{KravtsovScenario}.
For $\omega \ll \Delta_\xi$, the support of the two eigenfunctions are very similar and therefore the overlap function is essentially the same as the self-overlap,
given by the localization volume $\xi^d \times I_2$.
For $\omega$ of the order of the Fermi energy, $\omega\gg 2t$, one leaves the critical energy region where multifractality is relevant and enters a regime where overlaps happen on sites where one wave function can be treated perturbatively. This leads back to the quadratic behavior $C(\omega) \sim (t/\omega)^2$ described above. 

Let us come back to the nuclear polarization, which is approximately given by \eqref{POverlapLink}. The expression is particularly interesting for the case of small nuclear Zeeman energies $\omega_n< t$, where the stationary nuclear polarization senses the critical behavior of $C(\omega)$ in \eqref{KravtsovScenario}. Indeed we find
\begin{equation}
\frac{P_n}{P_n^{\tinytext{GOE}}}  \simeq C(\omega_n) \simeq
\left(\frac{t}{{\rm max}(\omega_n, \Delta_\xi)}\right)^\mu,
\end{equation}
which displays a distinct peak around $t \simeq t_c$, with a width $\delta t \simeq \omega_n^{1/d \nu}$.  In the limit of small $\omega_n$, i.e., for nuclear spins with tiny magnetic moments, the peak becomes increasingly sharp and high, the enhancement being a direct consequence of critical, multifractal scaling. 

A quantitative confirmation of the above scaling for the nuclear polarization is difficult numerically, since multifractality becomes visible only for large system sizes, which are beyond the scope of this work. Nevertheless, in the data shown in Fig. \ref{OVERLAP}, it is clearly visible that the maximum  in the overlap function $C(\omega_n)$  appears close to $t\simeq t_c$, and increases with decreasing  $\omega_n$.

\section{Conclusion}
In this work we have presented  a simplified approach to the dynamics of electron spins and the transfer of polarization to nuclear spins in the DNP
protocol. It is based on an approximate mapping to the free-fermion 3D Anderson model, for which we can obtain exact results 
regarding the driven steady state in large systems.
As the free fermion model is integrable, it fails to accurately describe the spin-temperature regime 
of an ergodic spin system. However, it exhibits a well-studied
Anderson transition, which allows us to investigate the influence of localization phenomena on the stationary nuclear polarization.
Remarkably, in a broad range of physical parameters relevant for DNP experiments, the optimal nuclear polarization is reached 
close to the localization transition. This agrees with the results obtained in \citei{deluca2016} for much smaller, but interacting systems.

 However, here we reach an additional  non-trivial prediction: the critical scaling close to the localization transition,  in particular the multifractal correlations of excitations close in energy, can further improve the efficiency of nuclear polarization, provided that the nuclear Zeeman energy $\omega_n$ is smaller than the hopping amplitude $t$. 
 We conjecture that a similar multifractal enhancement occurs in the real many-body problem of driven electronic spins. Instead of the density-density correlator, it is the local spin--spin correlator which is relevant there. Recent work ~\cite{serbyn2017} has indeed shown  that a similar power law enhancement arises at low frequencies when the overlap function is generalized to the appropriate matrix element of local operators.
This effect should come into play when the dipolar interaction $J$ between nearest neighbor electronic spins (the analogue of $t$ in the present work) exceeds the nuclear Zeeman energy $\omega_n$. 

 To be able to profit of such an enhancement at the many-body localization crossover, the dilution of electronic spins has to be tuned such that $w/J$ reaches its critical value (typically of order about $O(20)$ in $d=3$ dimensions), where $w=B \delta (g \mu_e)$ is the inhomogeneous broadening of the electronic Zeeman energy. Multifractal enhancement should then become relevant for ISS processes if the nuclei have magnetic moments smaller than $\delta (g \mu_e)/(w/J)_c$. 
  In practice this requires either fairly small nuclear moments, or a dipolar material with a large variation of electronic magnetic moments. The latter could be achieved e.g. by materials with a substantial local variation of Land\'e factors or strong internal fields. 
 
The above result suggests that DNP could in fact be used as a diagnostic 
tool to investigate localization physics, especially if it is probed by nuclear spins with small magnetic moments.

Our work opens several interesting directions: while here the dynamics of electron spins was frozen to their stationary state
due to negligible influence from the nuclei, it is interesting to explore different regimes, where the effect of the nuclei on 
the electron spins must be taken into account. In particular, one might consider that for sufficiently strong hyperfine interactions, 
an ergodic system of nuclei leads to the delocalization of the electron spins. As we mentioned, this situation can be approached within our framework by means of a Montecarlo simulation, which will be analyzed
in a forthcoming publication. 
However, opposite scenarii could arise as well, where strong hyperfine coupling creates electro-nuclear complexes that behave more classically and thus are more localized than the electronic system without hyperfine couplings.

Finally, in this work, we have assumed nuclear spins  to be a perfectly ergodic system, supposing that nuclear spin diffusion
occurs on a time scale that is much faster than all  other nuclear processes. Extending our approach to include a finite nuclear spin-diffusion time 
is a challenging direction, which requires a more sophisticated approach and will be addressed in future research.

\section*{Acknowledgments}
This work is supported by the
EPSRC  Quantum  Matter  in  and  out  of  Equilibrium grant
Ref.  EP/N01930X/1 (A.D.L.),  Investissements d'Avenir LabEx PALM
(ANR-10-LABX-0039-PALM), and ANR-16-CE30-0023-
01 (THERMOLOC).


\onecolumngrid
\appendix
\section{Diagonalization of the $3D$ - Anderson model \label{DiagHam}}
The starting point for the free fermions model for the electron spins is the Hamiltonian in Eq.~\eqref{dipham}, including the term that accounts for the dipolar interactions :
\begin{equation}
\HH{e}^{(dip)} = \sum_{i<j}\sum_{\ell=x,y,z}D_{ij} \hat{S}_i^\ell\hat{S}_j^\ell \,,
\end{equation}
which is analogous to the dipolar coupling between nuclear pairs of spins. The exact form of those dipolar couplings between two spins $\hat{S}_i,\hat{S}_j$ with respective gyromagnetic ratios $\gamma_i,\gamma_j$ - being $\gamma_e$ for an electron spin and $\gamma_n$ for a nuclear one - reads :
\begin{equation}
\label{dipcoupling}
D_{ij}=\frac{\mu_0 \gamma_1\gamma_2}{16\pi|\mathbf{r}_{ij}|^{3}} (1 - 3 \cos^2\theta_{ij})\;,
\end{equation}
with $\mathbf{r}_{ij}$ the vector connecting the two spins and $\theta_{ij}$ the angle that this vector forms with the external magnetic field. Note that the expression in Eq~\eqref{dipcoupling} varies strongly from pair to pair. 

At this point one performs the \emph{spin-to-fermion} transformation given in Eq.~\eqref{replSferm}, which results in a quadratic fermionic Hamiltonian 
\begin{equation}
\label{hamfermions}
\HH{\text{Ferm}}=\sum_{i,j}c^{\dagger}_i M_{ij} c_j \;, \text{with 
}M_{ij}=\omega_j \delta_{ij}+D_{ij}\;,
\end{equation}
after a constant shift of the energy and the Larmor frequencies $\omega_j$ of $\omega_e$ with respect to zero. Apart from dropping the Jordan-Wigner tails, we have neglected the interaction term arising from the dipolar coupling, and have retained  the hopping term only. 
We further retain a constant hopping term between nearest neighbors  assuming $D_{ij}=-t\delta_{i\pm1,j}$ in Eq.~\eqref{hamfermions}, which finally leads to the $3D$-Anderson model introduced in \eqref{hAnd}. Numerical diagonalization of the matrix $M_{ij}$ is performed and we denote $\phi_{\alpha i}$ as a single-particle eigen-function, i.e. an eigenvector of the matrix $M_{ij}$ with eigenvalue 
$\epsilon_\alpha$. We can then define creation/annihilation operators for those eigenstates, $a_{\alpha}=\sum_i \phi^*_{\alpha i}  c_i$, or equivalently $c_i=\sum_\alpha \phi_{\alpha i}a_\alpha$. The Hamiltonian \eqref{hamfermions} thus becomes
\begin{equation}
\HH{\text{ferm}}=\sum_{i,j} \sum_{\alpha\beta} \epsilon_\alpha c^{\dagger}_i 
\phi_{\alpha i}\delta_{\alpha \beta} \phi^*_{\beta j}  c_j 
=\sum_{\alpha} \epsilon_\alpha a^{\dagger}_\alpha  a_{\alpha}\;.
\end{equation}
The eigenfunctions are normalized and satisfy the orthogonality relations
$\sum_i \phi^*_{\alpha i} \phi_{\beta i}=\delta_{\alpha\beta}$.

\section{Microscopic derivation of the transition rates}
\subsection{Lattice and microwave-induced rates for the free fermions\label{mathRates}}
The explicit expressions for the transition rates induced by the reservoir and the microwaves can be obtained making several assumptions: We consider on the one hand that the lattice modes couple to local electronic spins only. This means that they can induce spin flips  respecting detailed balance at the temperature of the phonon bath $\beta^{-1}$. The rates  for a spin system were derived  in [\onlinecite{DeLuca2015}]. We can perform the \emph{spin-to-fermion} substitution in \eqref{replSferm}
\begin{align}
\label{bathEq}
W^{\text {bath}}_{\mu \to \nu}=&
\sum_{\substack{j=1\\ \ell=x,y}}^{N_e} 2\;\frac{h(\Delta E_{\mu \nu})}{T_{1e}}\;|\bra{\mu} \hat{S}_j^\ell \ket \nu|^2=
\sum_{\substack{j=1\\ \ell=+,-}}^{N_e} 2\;\frac{h(\Delta E_{\mu \nu})}{T_{1e}}\;|\bra{\mu} \hat{S}_j^\ell \ket \nu|^2
\quad \Longrightarrow \nonumber\\
W^{\text {bath}}_{\mu \to \nu}=&\sum_{j=1}^{N_e} \frac{h(\Delta E_{\mu \nu})}{T_{1e}}\; \left(|\bra{\mu}  c_j^\dagger   \ket \nu|^2 +
|\bra{\mu}  
c_j    \ket \nu|^2\right)
=\sum_{\alpha} \frac{h(\Delta E_{\mu \nu})}{T_{1e}}\; \left(|\bra{\mu}  c_\alpha^\dagger   \ket \nu|^2 +
|\bra{\mu}  
c_\alpha    \ket \nu|^2\right)\;.
\end{align} 
The function $h(\epsilon)=e^{\beta\epsilon}/(1+e^{\beta\epsilon})$ assures the Gibbs equilibrium at temperature $\beta^{-1}$ when the system is not irradiated; $T_{1e}$ is the typical time that the bath takes to induce a spin flip in the system and $\Delta E_{\mu\nu}=E_\mu-E_\nu$.

On the other hand, the system is being irradiated with the microwave field $\Hmw = \omega_1 \sum_j \hat{S}^x_j \cos(\omw t)$. This Hamiltonian is time-dependent, but one can perform the \emph{rotating-wave approximation} that neglects the fast-oscillating terms. In [\onlinecite{DeLuca2015}] the following rate for the microwave-induced transitions was obtained as :
 \begin{equation}
 \label{MWeq}
 W^{\text {MW}}_{\mu \to \nu}=
 \frac{4\omega_1^2 T_{2e}}{1+T_{2e}^2(|\Delta E_{\mu\nu}|-\omw)^2}\;|\bra{\mu} \sum_{j=1}^{N_e} \hat{S}_j^x \ket \nu|^2\;.
\end{equation}
Then one performs again the \emph{spin-to-fermion} substitution in equation \eqref{replSferm}.
\begin{equation}
 W^{\text {MW}}_{\mu \to \nu}=
\frac{\omega_1^2 T_{2e}}{1+T_{2e}^2(|\Delta E_{\mu\nu}|-\omw)^2} \;|\bra{\mu}\sum_{j=1}^{N_e}\left( c_j^\dagger + 
c_j  \right)  \ket \nu|^2
=\sum_{\alpha} \frac{T_{2e}\omega_1^2 |A_\alpha|^2 }{1+T_{2e}^2(|\Delta E_{\mu\nu}|-\omw)^2}  \;
|\bra{\mu} a_\alpha^\dagger + a_\alpha  \ket \nu|^2 \;,
 \end{equation}
with $A_\alpha = \sum_{i} \phi_{\alpha,i}$. Note that the rate of  bath induced transitions~\eqref{bathEq} is given by the sum of independent  single spin flips, unlike microwaves which act simultaneously on the total spin $\sum_j \hat S_j$. As a consequence, the microwave intensity $\omega_1$ is renormalized to $\omega_1|A_\alpha|$. In the absence of interactions it is easy to check that $|A_\alpha|=1$.  In the regime where \eqref{replSferm} is well justified, the hopping term $D_{ij} \ll \omega_j$, so that $A_\alpha \simeq 1$; for larger values of the hopping term, $A_\alpha$ exhibit unphysical fluctuations (although on average over $\alpha$ it remains true that $\langle A^2\;\rangle_\alpha = 1$): this is a manifestation of the breaking of the na\"ive spins-to-free fermions stated in \eqref{replSferm}. Consistently with our assumptions of weak dipolar coupling, we set from now on $|A_\alpha| = 1$.

\subsection{Hyperfine-induced rates: a single nuclear spin\label{hfMath}}
We are now interested in computing the transition rates induced by the presence of a single nuclear spin (labeled by the index $p$) weakly coupled to the electron spin at site $j$. We thus take the Hamiltonian in \eqref{hHF}, which induces on the one hand a nuclear spin flip $i_z\to\bar i_z$ and, on the other hand - due to the fact that expectation values of the local $\hat{S}_j^z$ operators are no longer conserved quantities - a change on the fermionic eigenstate. 
\begin{align}
W_{i_z\mu\to \bar i_z \nu}
=&\frac{T_{2n}}{1+T_{2n}^2\left(2i_z\omega_n+E_\mu-E_\nu\right)^2}\;|\bra{\mu, i_z} \HH{e--n}  \ket {\nu,\bar i_z}|^2
\nonumber\\
=&
\frac{T_{2n}A_{jp}^2}{1+T_{2n}^2\left(2i_z\omega_n+E_\mu-E_\nu\right)^2}\;|\bra{\mu,i_z} \hat{\mathcal I_x }  \left( c_j^\dagger c_j -1/2 \right) \ket {\nu,\bar i_z}|^2
\nonumber\\
=&
\frac{T_{2n}A_{jp}^2}{1+T_{2n}^2\left(2i_z\omega_n+E_\mu-E_\nu\right)^2}\;
|\bra{\mu}   \left( c_j^\dagger c_j -1/2 \right) \ket {\nu}|^2\;
|\bra{i_z} \hat{\mathcal I_x }  \ket {\bar i_z}|^2 
\nonumber\\
=&
\frac{T_{2n}A_{jp}^2/4}{1+T_{2n}^2\left(2i_z\omega_n+E_\mu-E_\nu\right)^2}\;|\bra{\mu} \left( c_j^\dagger c_j -1/2 \right) \ket \nu|^2\;,
\end{align}
The only matrix element that requires a little care to compute is:
\begin{align}
\bra{\mu} \left( c_j^\dagger c_j -\frac{1}{2} \right) \ket \nu =&
\sum_{\alpha\beta} \phi^*_{\alpha,j}\phi_{\beta,j} \bra{\mu} \left(  
a_\alpha^\dagger a_\beta -\frac{\delta_{\alpha,\beta}}{2} \right) \ket \nu  
=\nonumber \\
=&
\sum_{\alpha} \delta_{\mu,\nu} |\phi_{\alpha,j}|^2 \left( n_\alpha^\nu 
-\frac{1}{2} \right) 
+
\sum_{\alpha,\beta\neq\alpha} \phi^*_{\alpha,j} 
\phi_{\beta,j}(1-n_\alpha^\nu)n_\beta^\nu n_\alpha^\mu(1-n_\beta^\mu) 
\prod_{\gamma\neq\alpha,\beta}\delta_{n_\gamma^\nu n_\gamma^\mu}\;.
\end{align}
As we see, the total number of fermions is conserved. Two transitions are induced:
\begin{itemize}
\item A \emph{polarization-loss} transition which flips only the nuclear spin and leaves the fermionic eigenstate unchanged. The rate of this process is:
\begin{align}
\label{Wleakage1}
W^{\text{leak}}_{i_z\mu\to \bar i_z \mu}=
\frac{T_{2n}A_{jp}^2}{1+T_{2n}^2\omega_n^2} 
\sum_{\alpha, \beta}|\phi_{\alpha,j}|^2 |\phi_{\beta,j}|^2 
\left( n_\alpha^\mu -\frac{1}{2} \right) \left( n_\beta^\mu -\frac{1}{2} \right) \equiv
\frac{T_{2n}A_{jp}^2}{1+T_{2n}^2\omega_n^2} \left[ \sum_\alpha |\phi_{\alpha,j}|^2 \left( n_\alpha^\mu -\frac{1}{2} \right) \right]^2
 \;,
\end{align}
\item An \emph{ISS} transition which flips the nuclear spin exchanges the occupation number of two single particle fermionic modes (one being empty and the other full and vice-versa). The corresponding rate writes :
\begin{align}
\label{Wflipflop1}
W^{ISS}_{i_z\mu\to \bar i_z \nu\neq\mu}=
\frac{T_{2n}A_{jp}^2}{1+T_{2n}^2\left(2i_z\omega_n+E_\mu-E_\nu\right)^2
} 
\sum_{\alpha,\beta\neq\alpha}
|\phi_{\alpha,j}|^2 |\phi_{\beta,j}|^2 
(1-n_\alpha^\nu)n_\beta^\nu n_\alpha^\mu(1-n_\beta^\mu)
\prod_{\gamma\neq\alpha,\beta}\delta_{n_\gamma^\nu n_\gamma^\mu}\;.
\end{align}
\end{itemize}

\section{Derivation of the nuclear polarization in the absence of nuclear dipolar interactions
\label{singleMath}
}
For the sake of simplicity we restrict the hyperfine interactions to $A_{jp}=A_j\delta_{jp}$. Thus, in absence of nuclear dipolar interactions, a nucleus hyperpolarizes only in proximity of an electron spin $j$. Thus we can write the following master equation:
\begin{equation}
\label{mastergen}
\dot p^j_{i_z, \mu}= \sum_{\nu} \left( p^j_{\bar i_z, \nu} W^j_{\bar i_z\nu \to i_z\mu}- p^j_{i_z, \mu} 
W^j_{i_z \mu \to\bar i_z\nu}\right ) + \sum_{\nu} \left (p^j_{i_z, \nu} W^j_{i_z \nu \to i_z\mu}- p^j_{i_z, \mu} \;,
W^j_{i_z \mu \to i_z\nu} \right )\;,
\end{equation}
where $p^j_{i_z,\mu}$ is the probability of having the nucleus in proximity to the electron $j$ in the state $i_z$ and the fermions in the many-body eigenstate $\ket{\mu}$.  By summing over the possible electron states $\ket{\mu}$, we obtain the probability for the nucleus  to be in the state $i_z$: $p^j_{i_z} = \sum_\mu p^j_{i_z, \mu}$. The second sum in Eq.~\eqref{mastergen} contains only transitions which do not change the state of the nucleus and are eliminated by the sum over $\ket{\mu}$. Then we have:
\begin{equation}
\dot p^j_{i_z} = \sum_{\mu \nu}\left ( p^j_{\bar i_z, \nu} W^j_{\bar i_z\nu \to i_z\mu}- p^j_{i_z, \mu} 
W^j_{i_z \mu \to\bar i_z\nu} \right )= \sum_{\mu \nu}\left ( p^j_{\bar i_z, \mu} W^j_{\bar i_z\mu \to i_z\nu}- p^j_{i_z, \mu} 
W^j_{i_z \mu \to\bar i_z\nu}\right ) \;.
\end{equation}
To find the stationary state, we set $\dot p^j_{i_z} = 0$ and make the additional assumption 
-well justified for trityl electron radicals- that the dynamics of the nucleus does not affect the electron stationary state: this suggests the factorization $p^j_{i_z, \mu} = p^j_{i_z} \times p_{\mu}$, where $p_{\mu}$ is the probability of the fermions being in the many-body eigenstate $\ket\mu $. Let us define the transition rate $W^j_{i_z\to\bar i_z}=\sum_{\mu,\nu} p_\mu W^j_{i_z\mu\to\bar i_z \nu} \equiv  \Omega_j(2 i_z \omega_n)$, which is given by
\begin{align}
\scriptsize
\Omega_j(\omega) =
\sum_{\mu,\nu} \frac{A_{j}^2T_{2n} p_\mu}{1+T_{2n}^2\left(\omega+E_\mu-E_\nu\right)^2
} 
\sum_{\alpha, \beta}|\phi_{\alpha,j}|^2 |\phi_{\beta,j}|^2 
\left[ \delta_{\mu\nu}  \left( n_\alpha^\nu -\frac{1}{2} \right) \left( 
n_\beta^\nu -\frac{1}{2} \right) + 
(1-n_\alpha^\nu)n_\beta^\nu n_\alpha^\mu(1-n_\beta^\mu)
\prod_{\gamma\neq\alpha,\beta}\delta_{n_\gamma^\nu n_\gamma^\mu}\right]\;.
\end{align}
It is convenient to replace the single-particle occupation numbers of a given many-body eigenstate with the polarization of the original spin model, namely $P_\alpha^\mu= 2n_\alpha^\mu-1$ (which is either $\pm 1$) :
\begin{align}
\Omega_j(\omega) =
\frac{A_{j}^2T_{2n} }{4}
\sum_{\alpha, \beta}|\phi_{\alpha,j}|^2 |\phi_{\beta,j}|^2 
\sum_\mu p_\mu
\left[
\frac{(P^\mu_\alpha)^2 \delta_{\alpha,\beta}+P_\alpha^\mu P_\beta^\mu(1-\delta_{\alpha,\beta})}
{1+T_{2n}^2\omega^2}
+  \frac{
 (1+P_\alpha^\mu)(1-P_\beta^\mu)}{1+T_{2n}^2\left(\omega+\epsilon_\alpha-\epsilon_\beta\right)^2
} 
\right]\;.
\end{align}
The first term represents the polarization-loss and the second one represents the \emph{ISS} transition and involves the hopping of a fermion from the single-particle mode $\alpha$ to  the single-particle mode $\beta$ (two spin \emph{flip-flop}), thus the difference of energy reduces to $\Delta E_{\mu \nu}=\epsilon_\alpha-\epsilon_\beta$. In the stationary state $p_\mu\to p_\mu^{\text{st}}$, $\sum_\mu p_\mu^{\text{st}} P_\alpha^\mu=P_B(\epsilon_\alpha)$ and $\sum_\mu p_\mu^{\text{st}} P_\alpha^\mu P_\beta^\mu=P_B(\epsilon_\alpha)P_B(\epsilon_\beta)$ for $\alpha\neq\beta$, where $P_B(\epsilon_\alpha)$ is the Bloch polarization defined in equation~\eqref{pbloch}. We finally obtain :
\begin{align}
 \Omega_j(\omega) &= 
\frac{A_{j}^2T_{2n}}{4}
 \sum_{\alpha,\beta} 
 |\phi_{\alpha,j}|^2 |\phi_{\beta,j}|^2
\times \Bigg[
\frac{\delta_{\alpha,\beta}+P_B(\epsilon_\alpha) P_B(\epsilon_\beta)
(1-\delta_{\alpha,\beta})}{1+T_{2n}^2\omega^2} +
\frac{ (1+P_B(\epsilon_\alpha)) (1-P_B(\epsilon_\beta))
 (1-\delta_{\alpha,\beta})}{1+T_{2n}^2(\omega+\epsilon_\beta-\epsilon_\alpha)^2}
\Bigg]\;,
\end{align}
which has been simplified to yield Eq.~\eqref{OmegaFinal} by removing the vanishing contribution of the $\delta_{\alpha\beta}$ terms.

We obtain that the total rate for the transition of the $j$-th nucleus 
$\ket{\downarrow} \to \ket{\uparrow}$ reads
\begin{equation}
W_{\downarrow \to \uparrow}^{j}  \equiv \sum_{\mu,\nu} p_\mu^\text{\tiny stat}  W^{j}_{(\downarrow,\mu\to \uparrow ,\nu)} = \Omega_{j}(\omega_n)\;.
\end{equation}
Similarly, one obtains $W_{\downarrow \to \uparrow}^j = \Omega_j(-\omega_n)$.
In this way, we obtain the stationary polarization for the nucleus as:
\begin{equation}
P^j_n=
\frac{P_\uparrow-P_\downarrow}{P_\uparrow+P_\downarrow}
=\frac{\Omega_j(\omega_n) - \Omega_j(-\omega_n)}{\Omega_j(\omega_n) + \Omega_j(-\omega_n)}
= \frac{\sum\limits_{\alpha\beta} |\phi_{\alpha, j}|^2 |\phi_{\beta, j}|^2  \times \frac{P_\beta - P_\alpha}{1 + T_{2n}^2 (\epsilon_\beta - \epsilon_\alpha + \omega_n)^2}
}{
\sum\limits_{\alpha\beta} |\phi_{\alpha, j}|^2 |\phi_{\beta, j}|^2
\left(\frac{ P_\alpha P_\beta }{1 + T_{2n}^2 \omega_n^2
} + \frac{1 - P_\alpha P_\beta }{1 + T_{2n}^2 (\epsilon_\beta - \epsilon_\alpha + \omega_n)^2
}\right)}\;.
\end{equation}

\section{Derivation of the nuclear polarization for nuclei in the ETH phase\label{ETHMath}}
Here we want to compute the nuclear polarization of a strongly dipolar-interacting system of nuclear spins. In this case, the only conserved quantities are the total energy and the total magnetization of the nuclear spins. This corresponds to the regime of Eigenstate Thermalization Hypothesis (ETH). According to this hypothesis, we can assume that the matrix element of the local operator $\hat I^x_p$ between two nuclear eigenstates labeled by $\ket A$ and $\ket B$ takes the form
\begin{equation}
\label{ETHoffdiag}
\bra{A} \hat I^x_p \ket{B} = e^{-S(E_{AB})/2} f(E_{AB}, \Delta E_{AB}) R_{AB}
\end{equation}
where $E_{AB} = (E_A + E_B)/2$, $S(E)$ is the entropy and $R_{AB}$ contains all the fluctuations, 
being Gaussian random variables with zero average and unit variance. In the following, we set $R_{AB}^2= 1$ for simplicity.
As before, we now want now to compute the probability for the nuclear system to be at a given energy $E$. We define
\begin{equation}
 p(E) = e^{-S(E)} \sum_{A | E_A = E} p_A = e^{-S(E)} \sum_{A | E_A = E} \sum_\mu p_{A,\mu} 
\end{equation}
and again we make the assumption $p_{A,\mu} = p_A \; p_\mu$. We can perform a semi-classical integration that leads us to:
\begin{align}
\label{PEdot}
 \dot p(E) &= e^{-S(E)}\int dE' \sum_{\substack{A | E_A = E \\ B | E_B = E'}} \sum_{\mu,\nu} (p_{B} p_{\mu}  W_{(\mu,B) \to (\nu,A)}  - 
p_{A}p_{\mu}  W_{(\mu,A) \to (\nu,B)})=\nonumber \\
&=e^{-S(E)} \sum_{j} \int dE' \sum_{\substack{A | E_A = E \\ B | E_B = E'}} |\bra{A} I_x^q \ket{B}|^2 (p_{B} \Omega_j(-\Delta E_{AB})  - 
p_{A} \Omega_j(\Delta E_{AB}))=\nonumber\\
&=e^{-S(E)} \sum_{j} \int d\omega \sum_{\substack{A | E_A = E \\ B | E_B = E+\omega}} e^{-S(E + \omega/2)} f(E + \omega/2, \omega)^2 (p_{B} \Omega_j(-\omega)  - 
p_{A} \Omega_j(\omega))=\nonumber\\
&=\sum_{j} \int d\omega e^{S(E + \omega) -S(E + \omega/2)} f(E + \omega/2, \omega)^2 \left(p(E+\omega) \Omega_j(-\omega)  - 
p(E)\Omega_j(\omega)\right)
\end{align}
In order to fix the function $f(E, \omega)$, we compute the two-point correlation for an arbitrary observable $\hat O = \hat O_D+ \delta \hat O$ where $\hat O_D$ is the diagonal component. We have
\begin{multline}
\frac{1}{Z}\Tr[e^{-\beta H} \delta \hat O(t) \delta \hat O(0)] = \sum_{\mu,\nu} 
 e^{-\beta E_\mu } e^{i (E_\mu - E_\nu) t}\; |\bra{\mu} \delta \hat O \ket{\nu}|^2 =\\=
 \sum_{\mu,\nu} 
 e^{-\beta E_\mu } e^{i (E_\mu - E_\nu) t} e^{-S(E)} |f(E, \omega\bigr)|^2 R_{\mu \nu}^2 \simeq
 \int dE e^{-\beta E + \beta \omega/2 +  S(E)} \int d\omega |f(E, \omega)|^2 e^{i \omega t} 
\simeq \int d\omega |f(E_{\beta}, \omega)|^2 e^{(\beta/2 + it) \omega}  
\end{multline}
Applying this formula to the specific case $\delta \hat O = \hat I_x^q$ we have
\begin{equation}
 \int d\omega |f(E, \omega)|^2 e^{(\beta/2 + it) \omega}  = \frac{1}{Z} \Tr[ e^{-\beta_N(E) H_N} I_x^q(t) I_x^q]
\end{equation}
where $H_N$ is the Hamiltonian of the nuclear system (including the dipolar interactions), 
and $E$ is the average energy associated with the temperature $\beta_N(E)^{-1}$: $Z^{-1} \Tr[ e^{-\beta_N(E) \hat H_N} \hat H_N] = 
E$. In this way the function $f(E, \omega)$ is simply connected
to the Fourier-transform of the two-point correlation function of the $I_x^q$ operator. Note that such a correlation function can 
be accessed experimentally in the linear response regime.

As $E$ is a macroscopic energy, we can approximate $S(E + \omega/2) \simeq S(E) + S'(E) \omega/2$, with $S'(E) = \beta_N(E)$ and we arrive at
\begin{equation}
 \label{PEfinal}
  \dot p(E) =
  \sum_{j}  \int d\omega e^{\beta_N(E) \omega/2} |f(E, \omega)|^2 (p(E+\omega) \Omega_j(-\omega)  -  p(E)\Omega_j(\omega))
\end{equation}
This equation is valid for an arbitrary nuclear system under the hypothesis of ETH, i.e. Eq.~\eqref{ETHoffdiag} and only requires the 
knowledge of $f(E, \omega)$. 
If we now assume that the dipolar coupling between nuclear spins
is sufficient to establish an ETH, but negligible with respect to their
Zeeman gap $\omega_n$,  we can estimate $f(E, \omega)$
from the correlation function computed in absence of interactions
\begin{equation}
\label{nonintF}
 \frac{1}{Z} \Tr[ e^{-\beta_N(E) H_N} I_x^q(t) I_x^q] = \frac{\cosh((\beta_N(E) + i t)\omega_n/2)}{\cosh(\beta_N(E) \omega_n/2)} \quad \Rightarrow \quad
| f(E, \omega)|^2 =
 \frac{\delta (\omega - \omega_n) + \delta(\omega + \omega_n)}{2 \cosh(\beta_N(E) \omega_n/2)} 
 \end{equation}
which is peaked around the frequencies $\pm \omega_n$.
Here, we have associated to 
the microcanonical energy $E$ the corresponding canonical 
$\beta_N(E)$ using
\begin{equation}
 \label{totalEne}
E = \int dE' \; p(E') E' = \frac{N_n \omega_n \tanh(\beta_N(E) \omega_n/2)}{2}
\end{equation}
Injecting \eqref{nonintF} in \eqref{PEfinal}, we can now look for the stationary solution of \eqref{PEfinal} (i.e. $\dot p(E) = 0$), 
which requires
\begin{equation}
\frac{p(E+\omega_n)}{p(E)} = \frac{\sum_j \Omega_j(\omega_n)}{\sum_j \Omega_j(-\omega_n)} \quad \Rightarrow \quad 
P_n=\frac{\sum\limits_j \Omega_j(\omega_n) - \Omega_j(-\omega_n) }{\sum\limits_j \Omega_j(\omega_n) + \Omega_j(-\omega_n) }
\end{equation}
which we quoted in Eq.~\eqref{pnETH} of the main text. 

A simpler way to derive this result in this approximation would be to
assume that the eigenstates $\ket{A}$ of the nuclear Hamiltonian with a given number $N_+$ of spins up take the form
\begin{equation}
\ket{A} = \frac{1}{(\mathcal{N}_+)^{1/2}}\sum_{s = 1}^{\mathcal{N}_+} c_s^A \ket{s} \;, \qquad \mathcal{N}_+ = \binom{N_n}{N_+}
\end{equation}
where the sum over $s$ runs over all the $\mathcal{N}_+$ factorized configurations with $N_+$ spins up
and $c_s^A$ are random coefficients with zero average and variance $1$ and for simplicity we set $(c_s^A)^2 = 1$.
Then, one can explicitly compute the matrix element in \eqref{ETHoffdiag} which is non-vanishing only when
$N_A^+ = N_B^{+} \pm 1$. 
This leads to the master equation in the limit of large $N_n$:
\begin{equation}
\dot p(E) = \frac{N_+}{N_n}\sum_j \left[p(E + \omega_n) \Omega_j(\omega_n) - p(E) \Omega_j(-\omega_n)\right]
+ \left(1-\frac{N_+}{N_n} \right)\sum_j \left[p(E - \omega_n) \Omega_j(-\omega_n) - p(E) \Omega_j(\omega_n)\right]
\end{equation}
which is equivalent to \eqref{nonintF} and \eqref{PEfinal} once 
one uses that in equilibrium at temperature $\beta_N(E)^{-1}$, the ratio $N_+/N_n$ is $(1+\tanh(\beta_N(E) \omega_n/2))/2$.

\end{document}